\documentclass[12pt]{article}
\usepackage{amsmath,caption}
\usepackage[top=1.35in, bottom=1.35in, left=1.35in, right=1.35in]{geometry}

\DeclareCaptionStyle{italic}[justification=centering]
 {labelfont={bf},textfont={it},labelsep=colon}
\usepackage{graphicx,psfrag,epsf}
\usepackage{enumerate}
\usepackage{natbib}
\usepackage{url} 
\usepackage{booktabs, subfig, bm, paralist,mathpazo,tikz,todonotes,longtable,microtype} 
\usepackage[pdftex,colorlinks=true]{hyperref}
\definecolor{darkblue}{rgb}{0,0,.6}
\hypersetup{citecolor=darkblue,linkcolor=darkblue,urlcolor=darkblue}

\newcommand{\blind}{0}

\newcommand{\E}{\text{E}}

\newcommand{\F}{\mathcal{X}}
\graphicspath{{plots/}}
\DeclareMathOperator*{\argmin}{\arg\!\min}
\newsavebox\CBox
\def\textBF#1{\sbox\CBox{#1}\resizebox{\wd\CBox}{\ht\CBox}{\textbf{#1}}}

\definecolor{a0}{rgb}{0.0, 0.5, 0.0}
\definecolor{bistre}{rgb}{0.24, 0.17, 0.12}
\definecolor{amethyst}{rgb}{0.6, 0.4, 0.8}
\definecolor{blue-violet}{rgb}{0.54, 0.17, 0.89}
\definecolor{Rcolor}{RGB}{150,160,190}
\definecolor{blush}{rgb}{0.87, 0.36, 0.51}
\definecolor{brightturquoise}{rgb}{0.03, 0.91, 0.87}
\definecolor{burntorange}{rgb}{0.8, 0.33, 0.0}
\begin{document}

\def\spacingset#1{\renewcommand{\baselinestretch}%
{#1}\small\normalsize} \spacingset{1}

\if0\blind
{
  \title{\bf Grouped functional time series forecasting: \hbox{An application to age-specific mortality rates}}
  \author{Han Lin Shang\thanks{Han Lin Shang is a Senior Lecturer at the Research School of Finance, Actuarial Studies and Statistics, Australian National University, Canberra ACT 2601, Australia (E-mail: hanlin.shang@anu.edu.au). Rob J. Hyndman is a Professor at the Department of Econometrics and Business Statistics, Monash University, Melbourne VIC 3800, Australia (E-mail: rob.hyndman@monash.edu.au)}\hspace{.2cm}\\
    Research School of Finance, Actuarial Studies and Statistics \\
    Australian National University\\
    and \\
    Rob J Hyndman\\
    Department of Econometrics and Business Statistics \\
     Monash University}
  \maketitle
} \fi

\if1\blind
{
  \bigskip
  \bigskip
  \bigskip
  \begin{center}
    {\LARGE\bf Title}
\end{center}
  \medskip
} \fi

\bigskip
\begin{abstract}
Age-specific mortality rates are often disaggregated by different attributes, such as sex, state and ethnicity. Forecasting age-specific mortality rates at the national and sub-national levels plays an important role in developing social policy. However, independent forecasts at the sub-national levels may not add up to the forecasts at the national level. To address this issue, we consider reconciling forecasts of age-specific mortality rates, extending the methods of \citet{HAA+11} to functional time series, where age is considered as a continuum. The grouped functional time series methods are used to produce point forecasts of mortality rates that are aggregated appropriately across different disaggregation factors. For evaluating forecast uncertainty, we propose a bootstrap method for reconciling interval forecasts. Using the regional age-specific mortality rates in Japan, obtained from the Japanese Mortality Database, we investigate the one- to ten-step-ahead point and interval forecast accuracies between the independent and grouped functional time series forecasting methods. The proposed methods are shown to be useful for reconciling forecasts of age-specific mortality rates at the national and sub-national levels. They also enjoy improved forecast accuracy averaged over different disaggregation factors. Supplemental materials for the article are available online.
\end{abstract}

\vfill

\newpage
\spacingset{1.45} 

\section{Introduction}\label{sec:intro}

Functional time series often consist of random functions observed at regular time intervals. Depending on whether or not the continuum is also a time variable, functional time series can be grouped into two categories. On one hand, functional time series can arise by separating an almost continuous time record into natural consecutive intervals such as days, months or years \citep[see][]{HK12}. Examples include daily price curves of a financial stock \citep{KZ12}, and monthly sea surface temperature in climatology \citep{SH11}. On the other hand, functional time series can also arise when observations in a time period can be considered together as finite realizations of an underlying continuous function; for example, annual age-specific mortality rates in demography \citep[e.g.,][]{HU07,CM09}.

In either case, the functions  obtained form a time series $\{\F_t,~ t\in Z\}$, where each $\F_t$ is a (random) function $\F_t(z)$ and $z\in \mathcal{I}$ represents a continuum bounded within a finite interval. We refer to such data structures as functional time series. 

There has been a rapidly growing body of research on functional time series forecasting methods. From a parametric viewpoint, \cite{Bosq00} proposed the functional autoregressive (FAR) process of order 1 and derived one-step-ahead forecasts that are based on a regularized form of the Yule-Walker equations. \cite{KK16} proposed the functional moving average (FMA) process and introduce an innovations algorithm to obtain the best linear predictor. \cite{KKW16} proposed the FARMA process where a dimension reduction technique was used to reduce an infinite-dimensional object to a finite dimension, and then principal component scores can be modeled by vector autoregressive models. From a nonparametric perspective, \cite{BCS00} proposed functional kernel regression to measure the temporal dependence via a similarity measure characterized by neighborhood distance (also known as semi-metric), kernel function and bandwidth. From a semi-parametric viewpoint, \cite{AV08} put forward a semi-functional partial linear model that combines parametric and nonparametric models, and this semi-functional partial linear model allows us to consider additive covariates and to use a continuous path in the past to predict future values of a stochastic process.

Among many modeling techniques, functional principal component analysis (FPCA) has been used extensively for dimension reduction for a functional time series. As a data-driven basis function decomposition, FPCA can collapse an infinite-dimensional object to a finite dimension, without losing much information. \cite{HU07} use FPCA to decompose smoothed functional time series into a set of functional principal components and their associated principal component scores. The temporal dependency in the original functional time series is inherited by the correlation within each principal component score and the possible cross-correlations between principal component scores. \cite{HU07} applied univariate time series forecasting models to forecast these scores individually, while \cite{ANH15} considered a multivariate time series forecasting method to capture any correlations between principal component scores. Both univariate and multivariate time series forecasting methods have their own advantages and disadvantages \citep[see][for a comparison]{PS07, ANH15, Shang16}.

In this paper, we also use functional principal component regression as a forecasting technique, applied to a large multivariate set of functional time series with rich structure. There have been relatively few research contributions dealing with multivariate functional time series forecasting \citep[see for example,][]{CYC16,KMR16}. To our knowledge, there has been no study that takes account of aggregation constraints within multivariate functional time series forecasting. This is the gap we wish to address.

To be specific, we consider age-specific mortality rates observed annually as an example of a functional time series, where the continuum is the age variable. These age-specific mortality rates can be observed at the national level, and can be disaggregated by various attributes such as sex, state or ethnicity. Forecasts are often required for national mortality, as well as sub-national mortality disaggregated by different attributes. When a functional forecasting method is applied to each subset, the sum of the forecasts will not generally add up to the forecasts obtained by applying the method to the aggregated national data.

This problem is known as forecast reconciliation, which has been addressed for univariate time series forecasting. \cite{SW09} considered forecast reconciliation in the context of national account balancing, while \cite{HAA+11} demonstrated the usefulness of forecast reconciliation methods in the context of tourist demand.  In this paper, we develop reconciliation methods tailored for multivariate functional time series.

We put forward two statistical methods, namely bottom-up and optimal combination methods, to reconcile point and interval forecasts of age-specific mortality, and potentially improve the point and interval forecast accuracies. The bottom-up method involves forecasting each of the disaggregated series and then using simple aggregation to obtain forecasts for the aggregated series \citep{Kahn98}. This method works well where the bottom-level series have high signal-to-noise ratio. For highly disaggregated series, this does not tend to work well as the series become too noisy; also, any relationships between series are ignored. This motivated the development of an optimal combination method \citep{HAA+11}, where forecasts are obtained independently for all series at all levels of disaggregation and then a linear regression model is used with a generalized least-squares estimator to optimally combine and reconcile these forecasts. We propose a modification of this approach for use with functional time series.

Using the national and sub-national Japanese age-specific mortality rates from 1975 to 2013, we compare the point and interval forecast accuracies among the independent forecasting, bottom-up and optimal combination methods. For evaluating the point forecast accuracy, we consider the mean absolute forecast and root mean squared forecast errors, and found that the bottom-up method gives the most accurate overall point forecasts. For evaluating the interval forecast accuracy, we use the mean interval score, and again found that the bottom-up method gives the most accurate overall interval forecasts.

The rest of this paper is structured as follows. In Section~\ref{sec:2}, we describe the motivating data set, which is Japanese national and sub-national age-specific mortality rates. In Section~\ref{sec:3}, we describe the functional principal component regression for producing point and interval forecasts, then introduce grouped functional time series forecasting methods in Section~\ref{sec:4}. We evaluate and compare point and interval forecast accuracies between the independent and grouped functional time series forecasting methods in Sections~\ref{sec:5} and~\ref{sec:7}, respectively. Conclusions are presented in Section~\ref{sec:conclu}, along with some reflections on how the methods presented here can be further extended.

\section{Japanese age-specific mortality rates for 47 prefectures}\label{sec:2}

In many developed countries such as Japan, increases in longevity and an aging population have led to concerns regarding the sustainability of pensions, health and aged care systems \citep[see, for example,][]{Coulmas07, OECD13}. These concerns have resulted in a surge of interest amongst government policy makers and planners in accurately modeling and forecasting age-specific mortality rates. Sub-national forecasts of age-specific mortality rates are useful for informing policy within  local regions. Any improvement in the forecast accuracy of mortality rates will be beneficial for determining the allocation of current and future resources at the national and sub-national levels.

We study Japanese age-specific mortality rates from 1975 to 2013, obtained from the \citet{JMD15}. We consider ages from 0 to 99 in single years of age, while the last age group contains all ages at and beyond 100. The structure of the data is displayed in Table~\ref{tab:Japan} where each row denotes a level of disaggregation.  At the top level, we have total age-specific mortality rates for Japan. We can split these total mortality rates by sex, by region, or by prefecture. There are eight regions in Japan, which contain a total of 47 prefectures. The most disaggregated data arise when we consider the mortality rates for each combination of prefecture and sex, giving a total of $47\times2=94$ series. In total, across all levels of disaggregation, there are 168 series.

\begin{table}[!htbp]
\tabcolsep 0.27in
\centering
\caption{Hierarchy of Japanese mortality rates.}\label{tab:Japan}
\begin{tabular}{@{}lr@{}}
\toprule
Level                    & Number of series  \\
\midrule
Japan                    & 1   \\
Sex                      & 2   \\
Region                   & 8   \\
Sex $\times$ Region      & 16  \\
Prefecture               & 47  \\
Sex  $\times$ Prefecture & 94  \\
\midrule
Total                    & 168 \\
\bottomrule
\end{tabular}
\end{table}

\subsection{Rainbow plots}

\begin{figure}[!htbp]
\centering
\subfloat[Observed female mortality rates]
{\includegraphics[width=8.4cm]{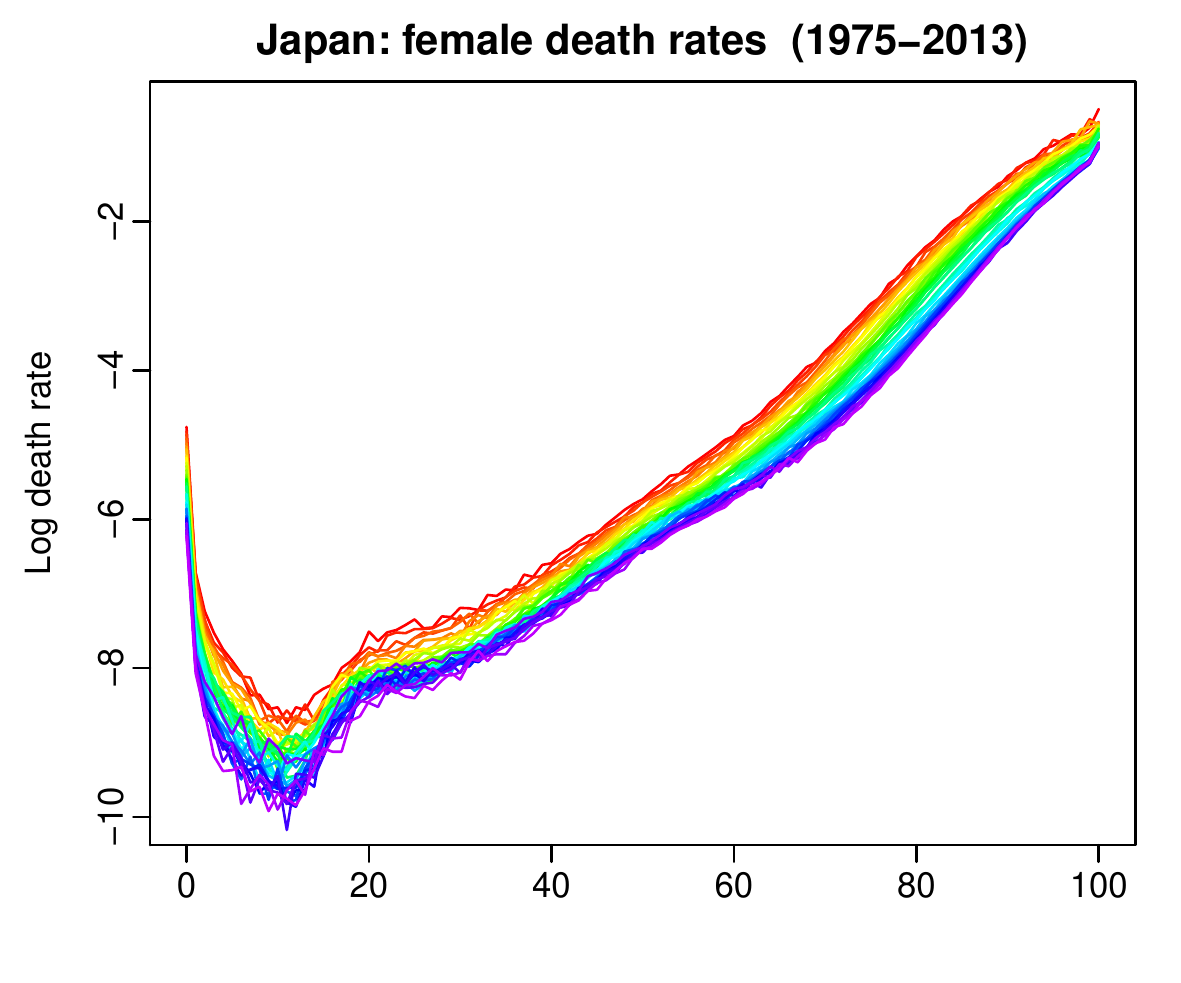}\label{fig:Japan_female_fig}}
\quad
\subfloat[Observed male mortality rates]
{\includegraphics[width=8.4cm]{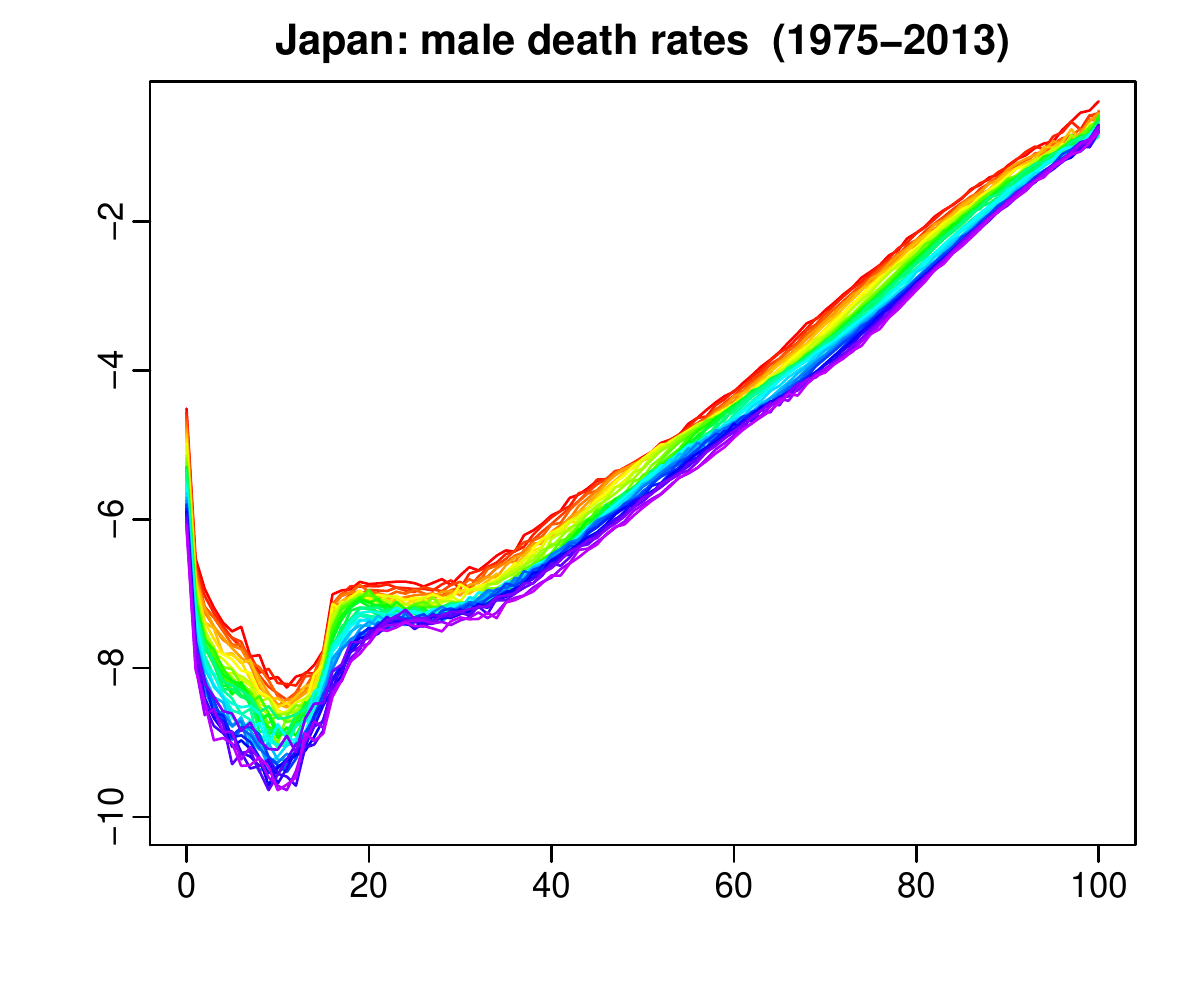}\label{fig:Japan_male_fig}}
\\
\subfloat[Smoothed female mortality rates]
{\includegraphics[width=8.4cm]{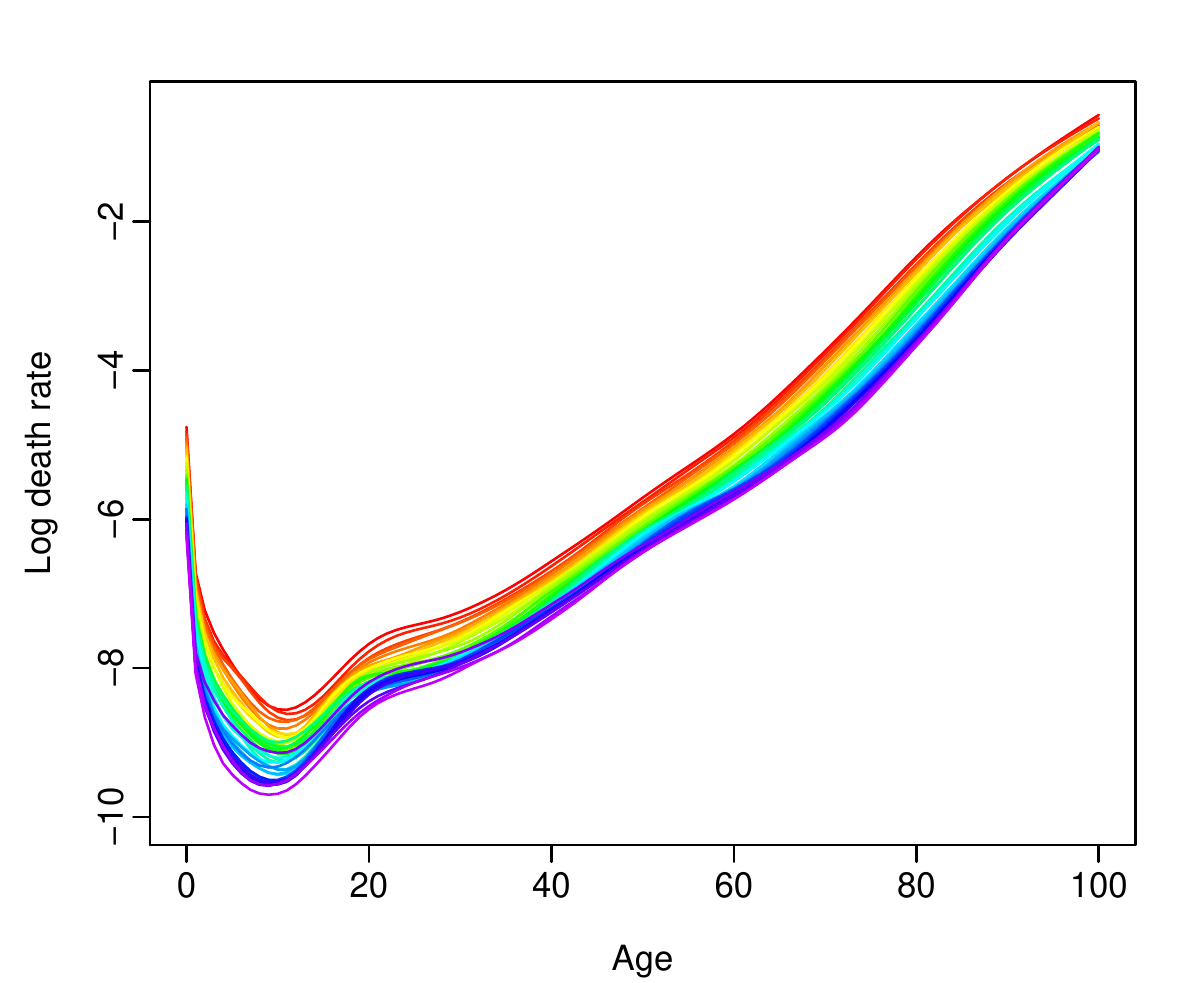}\label{fig:Japan_female_smooth_fig}}
\quad
\subfloat[Smoothed male mortality rates]
{\includegraphics[width=8.4cm]{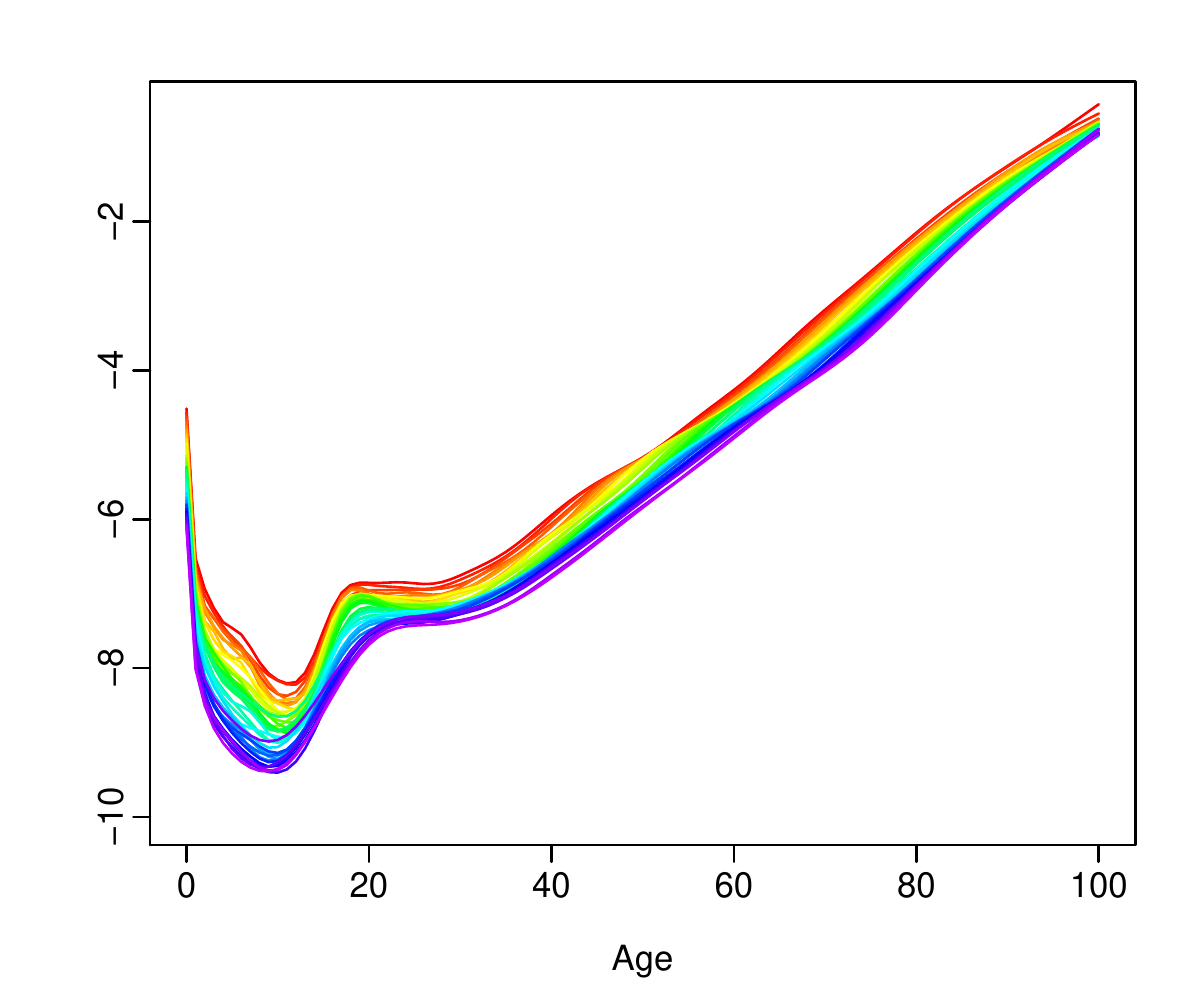}\label{fig:Japan_male_smooth_fig}}
\caption{Functional time series graphical displays}\label{fig:1}
\end{figure}

Figure~\ref{fig:1} shows rainbow plots of the female and male age-specific log mortality rates in Japan from 1975 to 2013 \citep{HS10}. The time ordering of the curves follows the color order of a rainbow, where curves from the distant past are shown in red and the more recent curves are shown in purple. The figures show typical mortality curves for a developed country, with rapidly decreasing mortality rates in the early years of life, followed by an increase during the teenage years, a plateau for young adults, and then a steady increase from about the age of 30. Females have lower mortality rates than males at all ages.

From Figures~\ref{fig:Japan_female_fig} and~\ref{fig:Japan_male_fig}, the observed mortality rates are not smooth across age due to observational noise. To obtain smooth functions and deal with possible missing values, we consider a penalized regression spline smoothing with monotonic constraint, described in Section~\ref{sec:3.2}. It takes into account the shape of log mortality curves \citep[see also][]{HU07, DPR11, Shang16b}.

Figures~\ref{fig:Japan_female_smooth_fig} and~\ref{fig:Japan_male_smooth_fig} demonstrate smooth age-specific mortality rates for Japanese females and males, and we apply smoothing to all series at different levels of disaggregation. We have developed a Shiny app \citep{Chang16} in R \citep{Team16} to allow interactive exploration of the smoothing of all the data; this is available in the online supplement.

From the rainbow plots in Figure~\ref{fig:1}, the age-specific mortality rates observed over years are not stationary, since the mean function changes over time. We implemented the stationarity test of \cite{HKR14} to confirm this, and found that mortality rates for both sexes in all prefectures were significantly non-stationary at the 5\% level (results not shown). 

\subsection{Image plots}

\begin{figure}[!htbp]
\centering
\includegraphics[width=\textwidth]{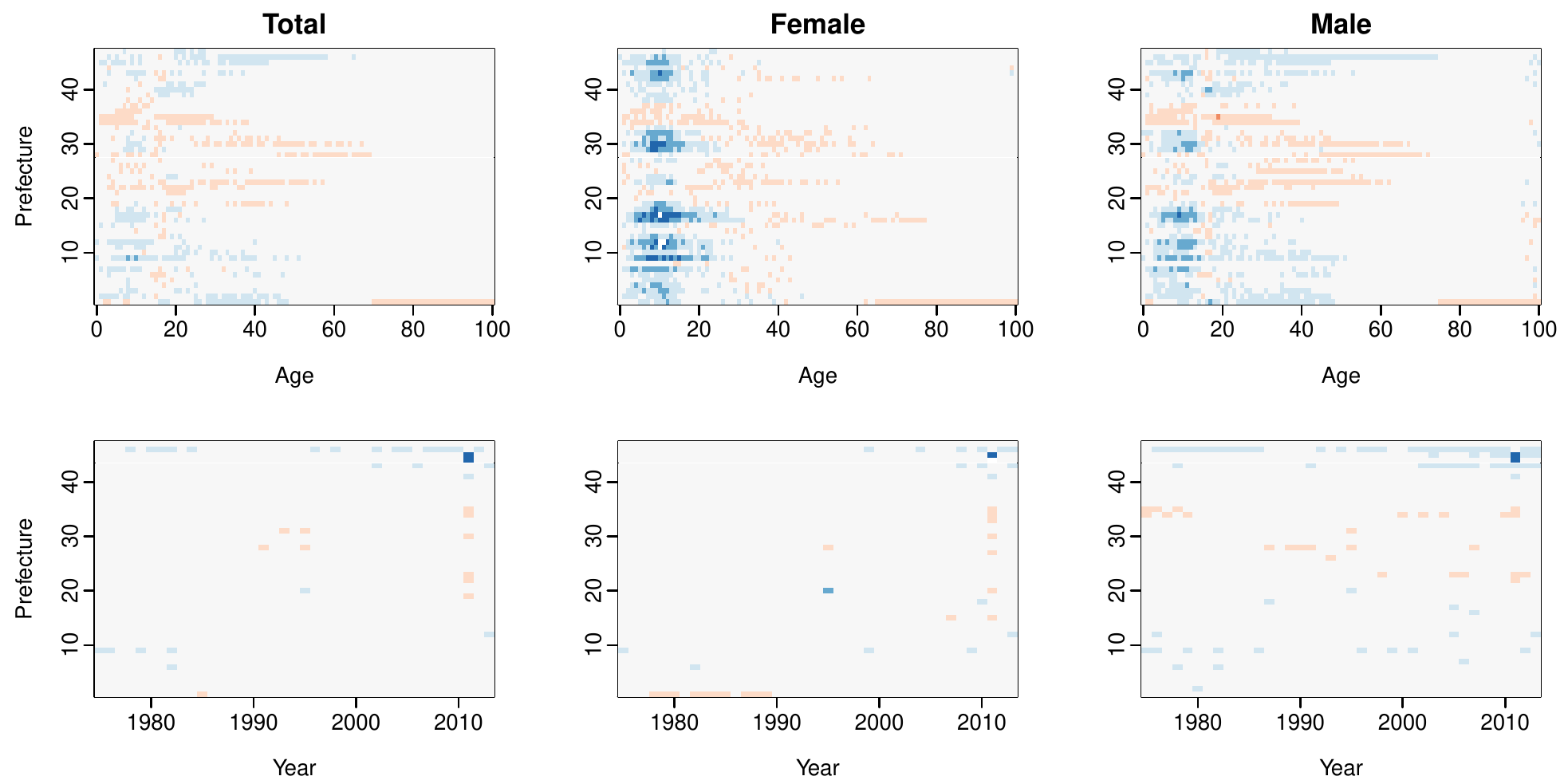}
\caption{Image plots showing log ratios of mortality rates. The top panel shows mortality rates averaged over years, while the bottom panel shows mortality rates averaged over ages. Prefectures are numbered geographically from north to south.}\label{fig:image}
\end{figure}

Another visual perspective of the data is shown in the image plots of Figure~\ref{fig:image}. Here we graph the log of the ratio of mortality rates for each prefecture to mortality rates for the whole country, thus allowing relative mortality comparisons to be made. A divergent color palette is used with blue representing positive values and orange denoting negative values. The prefectures are ordered geographically from north to south, so the most northerly prefecture (Hokkaid\={o}) is at the top of the panels, and the most southerly prefecture (Okinawa) is at the bottom.

The top row of panels shows mortality rates for each prefecture and age, averaged over all years. Several striking features become apparent. There are strong differences between the prefectures for children, especially females; this is possibly due to socio-economic differences, and accessibility of health services. The most southerly prefecture of Okinawa has particularly low mortality rates for older people; this is consistent with the extreme longevity for which Okinawa is famous \citep[see for example,][]{takata1987influence, suzuki2004successful, willcox2007aging}.

The bottom row of panels shows mortality rates for each prefecture and year, averaged over all ages. Here there is less information to be seen, but three outliers are highlighted. In 2011, in prefectures 44 (Miyagi) and 45 (Iwate) there was a large increase in mortality compared to other prefectures. These are northern coastal regions, and the inflated relative mortality is due to the tsunami of 11~March 2011. There is a corresponding decrease in relative mortality in some other provinces. In 1995, there is an increase in mortality for prefecture 20 (Hy\={o}go). This corresponds with the Kobe (Great Hanshin) earthquake of 17~January 1995.

Also evident is the decreased female mortality in Okinawa up until 1990, suggesting a recent decline in the relative mortality advantages enjoyed by residents in this region.

\section{Methodology}\label{sec:3}

\subsection{Functional principal component analysis}

Let $(\F_t: t\in Z)$ be an arbitrary functional time series. It is assumed that the observations $\F_t$ are elements of the Hilbert space $\mathcal{H}=L^2(\mathcal{I})$ equipped with the inner product $\langle w,v\rangle=\int_{\mathcal{I}} w(z)v(z)dz$, where $z$ represents a continuum and $\mathcal{I}$ represents the function support range. Each function is a square integrable function satisfying $\|\F_t\|^2=\int_{\mathcal{I}}\F_t^2(z)dz<\infty$ and associated norm. All random functions are defined on a common probability space $(\Omega, A, P)$. The notation $\F\in L_{\mathcal{H}}^p(\Omega, A, P)$ is used to indicate $\E(\|\F\|^p)<\infty$ for some $p>0$. When $p=1$, $\F(z)$ has the mean curve $\mu(z) = \E[\F(z)]$; when $p=2$, a non-negative definite covariance function is given by
\begin{equation}
c_{\F}(y, z) = \text{Cov}[\F(y),\F(z)] = \text{E}\left\{[\F(y) - \mu(y)][\F(z) - \mu(z)]\right\}\label{eq:covariance_fun}
\end{equation}
for all $y,z\in \mathcal{I}$. The covariance function $c_{\F}(y,z)$ in~\eqref{eq:covariance_fun} allows the covariance operator of $\F$, denoted by $\mathcal{K}_{\F}$ to be defined as
\begin{equation*}
\mathcal{K}_{\F}(\phi)(z) = \int_{\mathcal{I}} c_{\F}(y,z)\phi(y)dy.
\end{equation*}
Via Mercer's lemma, there exists an orthonormal sequence $(\phi_k)$ of continuous function in $L^2(\mathcal{I})$ and a non-increasing sequence $\lambda_k$ of positive number, such that
\begin{equation*}
c_{\F}(y,z) = \sum^{\infty}_{k=1}\lambda_k\phi_k(y)\phi_k(z), \qquad y,z\in \mathcal{I}.
\end{equation*}
By the separability of Hilbert spaces, the Karhunen--Lo\`{e}ve expansion of a stochastic process $\F(z)$ can be expressed as
\begin{align*}
\F(z) &= \mu(z) + \sum^{\infty}_{k=1}\beta_k \phi_k(z) \\
&=\mu(z)+\sum^{\infty}_{k=1}\sqrt{\lambda_k}\xi_k\phi_k(z),
\end{align*}
where $\xi_k = 1/\sqrt{\lambda_k}\int_{\mathcal{I}} [\F(z) - \mu(z)]\phi_k(z)dz$ is an uncorrelated random variable with a mean of zero and a unit variance. The principal component scores $\beta_k = \sqrt{\lambda_k}\xi_k$ are given by the projection of $\F(z) - \mu(z)$ in the direction of the $k$th eigenfunction $\phi_k$, i.e., $\beta_k = \langle\F(z) - \mu(z), \phi_k(z)\rangle$.

As a widely used dimension reduction technique, the FPCA summarizes the main features of an infinite-dimensional object by its finite key elements, and forms a base of functional principal component regression. For theoretical, methodological, and applied aspects of FPCA, consult the survey articles by \cite{Hall11}, \cite{Shang14}, \cite{WCM15} and \cite{RGS+16}.

When the function time series $(\F_t)$ is non-stationary, the principal components are not consistently estimated in this decomposition. However, the span of the basis functions is consistent \citep{Liebl2013-kg} and since our aim is to forecast linear combinations of the functions, this decomposition works even for non-stationary time series. In this way, our approach is similar to that of \citet{HU07,Lansangan2008-rr,Shen2009-hv} and others, all of whom use a functional principal components decomposition with non-stationary data.

\subsection{Nonparametric smoothing technique}\label{sec:3.2}

Functional data are intrinsically infinite dimensional, although we can only observe functional data at dense grid points \citep[see for example,][]{RS05} or sparse grid points \citep[see for example,][]{Muller05}. In practice, the observed data are often contaminated by random noise, referred to as measurement errors. As defined by \cite{WCM15}, measurement errors can be viewed as random fluctuations around a continuous and smooth function, or as actual errors in the measurement.

We assume that there are underlying $L_2$ continuous and smooth functions $\F_t(z)$ such that
\begin{equation*}
\mathcal{Y}_t(z_j) = \F_t(z_j) + \sigma_t(z_j)\varepsilon_{t,j}, \qquad t=1,\dots,n,\ j=1,\dots,p,
\end{equation*}
where $\mathcal{Y}_t(z_j)$ denotes the raw log mortality rates, $\{\varepsilon_{t,j}\}$ are independent and identically distributed (iid) random variables across $t$ and $j$ with a mean of zero and a unit variance, and $\sigma_t(z_j)$ allows for heteroskedasticity. We observe that measurement errors are realized only at those time points $z_{j}$ where measurements are being taken. As a result, these errors are treated as discretized data $\varepsilon_{t,j}$. However, in order to estimate the variance $\sigma_t^2(z_j)$, we assume that there is a latent smooth function $\sigma^2(z)$ observed at discrete time points.

Let $m_t(z_j)=\exp[\mathcal{Y}_t(z_j)]$ be the observed central mortality rates for age $z_j$ in year $t$ and define $E_t(z_j)$ to be the population of age $z_j$ at 30~June in year $t$ (often known as the ``exposure-at-risk''). The observed mortality rate follows a Poisson distribution with estimated variance
\begin{equation*}
\widehat{\sigma}^2_t(z_j) = \frac{1}{m_t(z_j) E_t(z_j)}.
\end{equation*}

For modeling age-specific log mortality, \cite{HU07} advocated the application of  weighted penalized regression splines with a monotonic constraint for ages above 65, where the weights are equal to the inverse variances, $w_t(z_j) = 1/\widehat{\sigma}_t^2(z_j)$. For each year $t$,
\begin{equation*}
\widehat{\F}_t(z_j) = \argmin_{\theta_t(z_j)}\sum^M_{j=1}w_t(z_j)\left|\mathcal{Y}_{t}(z_j)-\theta_t(z_j)\right|+\lambda \sum^{M-1}_{j=1}\left|\theta^{'}_{t}(z_{j+1})-\theta^{'}_t(z_j)\right|,
\end{equation*}
where $z_j$ represents different ages (grid points) in a total of $M$ grid points, $\lambda$ represents a smoothing parameter, $\theta^{'}$ denotes the first derivative of smooth function $\theta$, which can be both approximated by a set of $B$-splines \citep[see for example,][]{deBoor01}. The $L_1$ loss function and $L_1$ penalty function are used to obtain robust estimates. This monotonic constraint helps to reduce the noise from estimation of high ages \citep[see also][]{DPR11}.

\subsection{Functional principal component regression}

By using FPCA, a time series of smoothed functions $\bm{\F}(z) = \{\F_1(z),\dots,\F_n(z)\}$ is decomposed into orthogonal functional principal components and their associated principal component scores, given by
\begin{align}
\F_t(z) &= \mu(z) + \sum^{\infty}_{k=1}\beta_{t,k}\phi_k(z) \notag\\
&= \mu(z) + \sum^K_{k=1}\beta_{t,k}\phi_k(z) + e_t(z), \label{eq:fpca}
\end{align}
where $\mu(z)$ is the mean function; $\{\phi_1(z),\dots,\phi_K(z)\}$ is a set of the first $K$ functional principal components; $\bm{\beta}_1 = (\beta_{1,1},\dots,\beta_{1,n})^{\top}$ and $\left\{\bm{\beta}_{1},\dots,\bm{\beta}_K\right\}$ denotes a set of principal component scores and $\bm{\beta}_k\sim N(0,\lambda_k)$ where $\lambda_k$ is the $k$th eigenvalue of the covariance function in~\eqref{eq:covariance_fun}; $e_t(z)$ denotes the model truncation error function with a mean of zero and a finite variance; and $K<n$ is the number of retained components. Expansion~\eqref{eq:fpca} facilitates dimension reduction as the first $K$ terms often provide a good approximation to the infinite sums, and thus the information contained in $\bm{\F}(z)$ can be adequately summarized by the $K$-dimensional vector $(\bm{\beta}_1,\dots,\bm{\beta}_K)$.

Although it can be a research topic on its own, there are several approaches for selecting $K$:
\begin{inparaenum}
\item[(1)] scree plots or the fraction of variance explained by the first few functional principal components \citep{Chiou12};
\item[(2)] pseudo-versions of Akaike information criterion and Bayesian information criterion \citep{YMW05};
\item[(3)] predictive cross validation leaving out one or more curves \citep{RS91};
\item[(4)] bootstrap methods \citep{HV06}.
\end{inparaenum}
Here, the value of $K$ is chosen as the minimum that reaches a certain level of the proportion of total variance explained by the $K$ leading components such that
\begin{equation*}
K=\argmin_{K:K\geq 1}\left\{\sum^K_{k=1}\widehat{\lambda}_k\Big/\sum^{\infty}_{k=1}\widehat{\lambda}_k\mathbb{1}_{\left\{\widehat{\lambda}_k>0\right\}}\geq \delta \right\},
\end{equation*}
where $\delta = 90\%$, $\mathbb{1}_{\left\{\widehat{\lambda}_k>0\right\}}$ is to exclude possible zero eigenvalues, and $\mathbb{1}\{\cdot\}$ represents the binary indicator function.

In a dense and regularly spaced functional time series, the mean function $\widehat{\mu}(z) = \frac{1}{n}\sum^n_{t=1}\F_t(z)$ and covariance function $\widehat{c}_{\F}(y,z)$ can be empirically estimated and they are shown to be consistent under the weak dependency \citep{HK10}. From the empirical covariance function, we can extract empirical functional principal component functions $\bm{\mathcal{B}} = \left\{\widehat{\phi}_1(z),\dots,\widehat{\phi}_K(z)\right\}$ using singular value decomposition. Conditioning on the smoothed functions $\bm{\F}(z)=\{\F_1(z),\dots,\F_n(z)\}$ and the estimated functional principal components $\bm{\mathcal{B}}$, the $h$-step-ahead point forecast of $\F_{n+h}(z)$ can be obtained as
\begin{equation*}
\widehat{\F}_{n+h|n}(z) = \text{E}[\F_{n+h}(z)|\bm{\F}(z),\bm{\mathcal{B}}] = \widehat{\mu}(z) + \sum^K_{k=1}\widehat{\beta}_{n+h|n,k}\widehat{\phi}_k(z),
\end{equation*}
where $\widehat{\beta}_{n+h|n,k}$ represents the time series forecasts of the $k$th principal component scores, which can be obtained by using a univariate time series forecasting method, which can handle non-stationarity of the principal component scores.

\subsection{A univariate time series forecasting method}\label{sec:3.4}

\cite{HS09} considered a univariate time series forecasting method to obtain $\widehat{\beta}_{n+h|n,k}$, such as autoregressive integrated moving average (ARIMA) model. This univariate time series forecasting method is able to model non-stationary time series containing a stochastic trend component. Since the yearly age-specific mortality rates do not contain seasonality, the ARIMA has a general form of
\begin{equation*}
(1-\psi_1B-\cdots-\psi_pB^p)(1-B)^d\bm{\beta}_k = \alpha + (1+\theta_1B+\cdots+\theta_qB^q)\bm{w}_k,
\end{equation*}
where $\alpha$ represents the intercept, $(\psi_1,\dots,\psi_p)$ denote the coefficients associated with the autoregressive component, $(\theta_1,\dots,\theta_q)$ denote the coefficients associated with the moving average component, $B$ denotes the backshift operator, $d$ denotes the differencing operator, and $\bm{w}_k = \left\{w_{1,k},\dots,w_{n,k}\right\}$ represents a white-noise error term. We use the automatic algorithm of \cite{HK08} to choose the optimal orders of autoregressive $p$, moving average $q$ and difference order $d$. The value of $d$ is selected based on successive Kwiatkowski-Phillips-Schmidt-Shin (KPSS) unit-root tests \citep{KPSS92}. KPSS tests are used for testing the null hypothesis that an observable time series is stationary around a deterministic trend. We first test the original time series for a unit root; if the test result is significant, then we test the differenced time series for a unit root. The procedure continues until we obtain our first insignificant result. Having determined $d$, the orders of $p$ and $q$ are selected based on the optimal Akaike information criterion (AIC) with a correction for small sample sizes \citep{Akaike74,Hurvich1989-hm}. Having identified the optimal ARIMA model, maximum likelihood method can then be used to estimate the parameters.

\section{Grouped functional time series forecasting techniques}\label{sec:4}

\subsection{Notation}

\begin{figure}[!htbp]
\centering\begin{tikzpicture}
\tikzstyle{every node}=[minimum size = 8mm]
\tikzstyle[level distance=10cm] \tikzstyle[sibling distance=40cm]
\tikzstyle{level 3}=[sibling distance=16mm,font=\footnotesize]
\tikzstyle{level 2}=[sibling distance=22mm,font=\small]
\tikzstyle{level 1}=[sibling distance=40mm,font=\normalsize]
\node[circle,draw]{Japan}
   child {node[circle,draw] {R1}
   	     child {node[circle,draw] {P1}}}
   child {node[circle,draw] {R2}
   		child {node[circle,draw] {P2}}
      child {node {$\cdots$}edge from parent[draw=none]}
		child {node[circle,draw] {P7}}
		}
  child {node {$\cdots$}edge from parent[draw=none]}
   child {node[circle,draw] {R8}
   		child{node[circle,draw] {P40}}
	     child{node {$\cdots$}edge from parent[draw=none]}
  	child{node[circle,draw] {P47}}
 };
\end{tikzpicture}
\caption{The Japanese geographical hierarchy tree diagram, with eight regions and 47 prefectures.}\label{fig:2}
\end{figure}
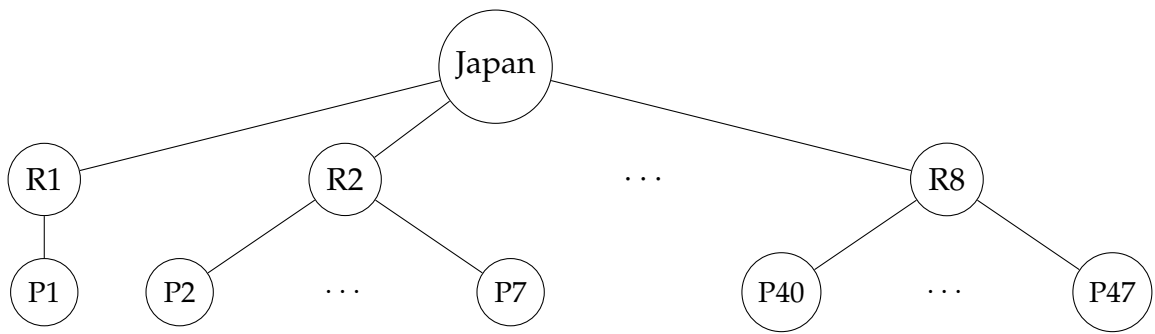

For ease of explanation, we will introduce the notation using the Japanese example. The generalization to other contexts should be apparent. The Japanese data follow a multi-level geographical hierarchy coupled with a sex grouping variable. The hierarchy is shown in Figure~\ref{fig:2}. Japan is split into eight regions, which in turn can be split into 47 prefectures.

The data can also be split by sex. So each of the nodes in the geographical hierarchy can also be split into both males and females. We refer to a particular disaggregated series using the notation X*S meaning the geographical area X and the sex S, where X can take the values shown in Figure~\ref{fig:2} and S can take values M (males), F (females) or T (total). For example: R1*F denotes females in Region 1; P1*T denotes females and males in Prefecture 1; Japan*M denotes males in Japan; and so on.

Let $E_{\text{X*S},t}(z)$ denote the exposure-at-risk for series X*S in year $t$ and age $z$, and let $D_{\text{X*S},t}(z)$ be the number of deaths for series X*S in year $t$ and age $z$. Then the age-specific mortality rate is given by $R_{\text{X*S},t}(z) = D_{\text{X*S},t}(z)/E_{\text{X*S},t}(z)$. To simplify expressions, we will drop the age argument $(z)$. Then for a given age, we can write

\begin{footnotesize}
\arraycolsep=0.1cm
\[
\underbrace{ \left[
\begin{array}{l}
R_{\text{Japan*T},t} \\
R_{\textcolor{red}{\text{Japan*F},t}} \\
R_{\textcolor{red}{\text{Japan*M},t}} \\
R_{\textcolor{a0}{\text{R1*T},t}} \\
R_{\textcolor{a0}{\text{R2*T},t}} \\
\vdots \\
R_{\textcolor{a0}{\text{R8*T},t}} \\
R_{\textcolor{blue-violet}{\text{R1*F},t}} \\
R_{\textcolor{blue-violet}{\text{R2*F},t}} \\
\vdots \\
R_{\textcolor{blue-violet}{\text{R8*F},t}} \\
R_{\textcolor{burntorange}{\text{R1*M},t}} \\
R_{\textcolor{burntorange}{\text{R2*M},t}} \\
\vdots \\
R_{\textcolor{burntorange}{\text{R8*M},t}} \\
R_{\textcolor{blue}{\text{P1*T},t}} \\
R_{\textcolor{blue}{\text{P2*T},t}} \\
\vdots \\
R_{\textcolor{blue}{\text{P47*T},t}} \\
R_{\textcolor{purple}{\text{P1*F},t}} \\
R_{\textcolor{purple}{\text{P1*M},t}} \\
R_{\textcolor{purple}{\text{P2*F},t}} \\
R_{\textcolor{purple}{\text{P2*M},t}} \\
\vdots \\
R_{\textcolor{purple}{\text{P47*F},t}} \\
R_{\textcolor{purple}{\text{P47*M},t}} \\ \end{array}
\right]}_{\bm{R}_t} =
\underbrace{\left[
\begin{array}{ccccccccccc}
\frac{E_{\text{P1*F},t}}{E_{\text{Japan*T},t}} & \frac{E_{\text{P1*M},t}}{E_{\text{Japan*T},t}} & \frac{E_{\text{P2*F},t}}{E_{\text{Japan*T},t}} & \frac{E_{\text{P2*M},t}}{E_{\text{Japan*T},t}}  & \frac{E_{\text{P3*F},t}}{E_{\text{Japan*T},t}} & \frac{E_{\text{P3*M},t}}{E_{\text{Japan*T},t}} & \cdots & \frac{E_{\text{P47*F},t}}{E_{\text{Japan*T},t}} & \frac{E_{\text{P47*M},t}}{E_{\text{Japan*T},t}} \\
\textcolor{red}{\frac{E_{\text{P1*F},t}}{E_{\text{Japan*F},t}}} & \textcolor{red}{0} & \textcolor{red}{\frac{E_{\text{P2*F},t}}{E_{\text{Japan*F},t}}} & \textcolor{red}{0} & \textcolor{red}{\frac{E_{\text{P3*F},t}}{E_{\text{Japan*F},t}}} & \textcolor{red}{0} & \cdots & \textcolor{red}{\frac{E_{\text{P47*F},t}}{E_{\text{Japan*F},t}}} & \textcolor{red}{0} \\
\textcolor{red}{0} & \textcolor{red}{\frac{E_{\text{P1*M},t}}{E_{\text{Japan*M},t}}}  & \textcolor{red}{0} & \textcolor{red}{\frac{E_{\text{P2*M},t}}{E_{\text{Japan*M},t}}} & \textcolor{red}{0} & \textcolor{red}{\frac{E_{\text{P3*M},t}}{E_{\text{Japan*M},t}}} & \cdots & \textcolor{red}{0} & \textcolor{red}{\frac{E_{\text{P47*M},t}}{E_{\text{Japan*M},t}}} \\
\textcolor{a0}{\frac{E_{\text{P1*F},t}}{E_{\text{R1,T},t}}} & \textcolor{a0}{\frac{E_{\text{P1*M},t}}{E_{\text{R1,T},t}}} & \textcolor{a0}{0} & \textcolor{a0}{0} & \textcolor{a0}{0} & \textcolor{a0}{0} & \cdots  & \textcolor{a0}{0} & \textcolor{a0}{0} \\
\textcolor{a0}{0} & \textcolor{a0}{0} & \textcolor{a0}{\frac{E_{\text{P2*F},t}}{E_{\text{R2,T},t}}} & \textcolor{a0}{\frac{E_{\text{P2*M},t}}{E_{\text{R2,T},t}}} & \textcolor{a0}{\frac{E_{\text{P3*F},t}}{E_{\text{R2,T},t}}} & \textcolor{a0}{\frac{E_{\text{P3*M},t}}{E_{\text{R2,T},t}}} & \cdots & \textcolor{a0}{0} & \textcolor{a0}{0} \\
\vdots & \vdots & \vdots & \vdots & \vdots & \vdots & \cdots & \vdots & \vdots \\
\textcolor{a0}{0} & \textcolor{a0}{0} & \textcolor{a0}{0} & \textcolor{a0}{0} & \textcolor{a0}{0} & \textcolor{a0}{0} & \cdots & \textcolor{a0}{\frac{E_{\text{P47*F},t}}{E_{\text{R8,T},t}}} & \textcolor{a0}{\frac{E_{\text{P47*M},t}}{E_{\text{R8,T},t}}} \\
\textcolor{blue-violet}{\frac{E_{\text{P1*F},t}}{E_{\text{R1,F},t}}} & \textcolor{blue-violet}{0} & \textcolor{blue-violet}{0} & \textcolor{blue-violet}{0} & \textcolor{blue-violet}{0} & \textcolor{blue-violet}{0} &  \cdots & \textcolor{blue-violet}{0} & \textcolor{blue-violet}{0} \\
\textcolor{blue-violet}{0} & \textcolor{blue-violet}{0} & \textcolor{blue-violet}{\frac{E_{\text{P2*F},t}}{E_{\text{R2,F},t}}} & \textcolor{blue-violet}{0} & \textcolor{blue-violet}{\frac{E_{\text{P3*F},t}}{E_{\text{R2,F},t}}} & \textcolor{blue-violet}{0} & \cdots & \textcolor{blue-violet}{0} & \textcolor{blue-violet}{0}  \\
\vdots & \vdots & \vdots & \vdots & \vdots & \vdots & \cdots & \vdots & \vdots \\
\textcolor{blue-violet}{0} & \textcolor{blue-violet}{0}  & \textcolor{blue-violet}{0}  & \textcolor{blue-violet}{0}  & \textcolor{blue-violet}{0}  & \textcolor{blue-violet}{0}  & \cdots & \textcolor{blue-violet}{\frac{E_{\text{P47*F},t}}{E_{\text{R8,F},t}}} & \textcolor{blue-violet}{0}\\
\textcolor{burntorange}{0} & \textcolor{burntorange}{\frac{E_{\text{P1*M},t}}{E_{\text{R1,M},t}}} & \textcolor{burntorange}{0} &\textcolor{burntorange}{0} & \textcolor{burntorange}{0} & \textcolor{burntorange}{0} & \cdots & \textcolor{burntorange}{0} & \textcolor{burntorange}{0} \\
\textcolor{burntorange}{0} & \textcolor{burntorange}{0} & \textcolor{burntorange}{0} & \textcolor{burntorange}{\frac{E_{\text{P2*M},t}}{E_{\text{R2,M},t}}} & \textcolor{burntorange}{0} & \textcolor{burntorange}{\frac{E_{\text{P3*M},t}}{E_{\text{R2,M},t}}} & \cdots & \textcolor{burntorange}{0} & \textcolor{burntorange}{0} \\
\vdots & \vdots & \vdots & \vdots & \vdots & \vdots & \cdots & \vdots & \vdots \\
\textcolor{burntorange}{0} & \textcolor{burntorange}{0} & \textcolor{burntorange}{0} & \textcolor{burntorange}{0} & \textcolor{burntorange}{0} & \textcolor{burntorange}{0} & \cdots & \textcolor{burntorange}{0} & \textcolor{burntorange}{\frac{E_{\text{P47*M},t}}{E_{\text{R8,M},t}}} \\
\textcolor{blue}{\frac{E_{\text{P1*F},t}}{E_{\text{P1,T},t}}} & \textcolor{blue}{\frac{E_{\text{P1*M},t}}{E_{\text{P1,T},t}}} & \textcolor{blue}{0} & \textcolor{blue}{0} & \textcolor{blue}{0} & \textcolor{blue}{0} & \cdots & \textcolor{blue}{0} & \textcolor{blue}{0} \\
\textcolor{blue}{0} & \textcolor{blue}{0}  &  \textcolor{blue}{\frac{E_{\text{P2*F},t}}{E_{\text{P2,T},t}}} & \textcolor{blue}{\frac{E_{\text{P2*M},t}}{E_{\text{P2,T},t}}} & \textcolor{blue}{0} & \textcolor{blue}{0}  & \cdots & \textcolor{blue}{0} & \textcolor{blue}{0} \\
\vdots & \vdots & \vdots & \vdots & \vdots & \vdots & \cdots & \vdots & \vdots \\
\textcolor{blue}{0} & \textcolor{blue}{0} & \textcolor{blue}{0} & \textcolor{blue}{0} & \textcolor{blue}{0} & \textcolor{blue}{0} & \cdots & \textcolor{blue}{\frac{E_{\text{P47*F},t}}{E_{\text{P47,T},t}}} & \textcolor{blue}{\frac{E_{\text{P47*M},t}}{E_{\text{P47,T},t}}} \\
\textcolor{purple}{1} & \textcolor{purple}{0} & \textcolor{purple}{0} & \textcolor{purple}{0} & \textcolor{purple}{0} & \textcolor{purple}{0} & \cdots & \textcolor{purple}{0} & \textcolor{purple}{0} \\
\textcolor{purple}{0} & \textcolor{purple}{1} & \textcolor{purple}{0} & \textcolor{purple}{0} & \textcolor{purple}{0} & \textcolor{purple}{0} & \cdots & \textcolor{purple}{0} & \textcolor{purple}{0} \\
\textcolor{purple}{0} & \textcolor{purple}{0} & \textcolor{purple}{1} & \textcolor{purple}{0} & \textcolor{purple}{0} & \textcolor{purple}{0} & \cdots & \textcolor{purple}{0} & \textcolor{purple}{0} \\
\textcolor{purple}{0} & \textcolor{purple}{0} & \textcolor{purple}{0} & \textcolor{purple}{1} & \textcolor{purple}{0} & \textcolor{purple}{0} & \cdots & \textcolor{purple}{0} & \textcolor{purple}{0} \\
\vdots & \vdots & \vdots & \vdots & \vdots & \vdots & \cdots & \vdots & \vdots  \\
\textcolor{purple}{0} & \textcolor{purple}{0} & \textcolor{purple}{0} & \textcolor{purple}{0} & \textcolor{purple}{0} & \textcolor{purple}{0} & \cdots  & \textcolor{purple}{1} & \textcolor{purple}{0}\\
\textcolor{purple}{0} & \textcolor{purple}{0} & \textcolor{purple}{0} & \textcolor{purple}{0} & \textcolor{purple}{0} & \textcolor{purple}{0} & \cdots & \textcolor{purple}{0} & \textcolor{purple}{1} \\
\end{array}
\right]}_{\bm{S}_t}
\underbrace{\left[
\begin{array}{l}
R_{\text{P1*F},t} \\
R_{\text{P1*M},t} \\
R_{\text{P2*F},t} \\
R_{\text{P2*M},t} \\
\vdots \\
R_{\text{P47*F},t} \\
R_{\text{P47*M},t} \\
\end{array}
\right]}_{\bm{b}_t}
\]
\end{footnotesize}
or $\bm{R}_t = \bm{S}_t \bm{b}_t$ where $\bm{R}_t$ is a vector containing all series at all levels of disaggregation, $\bm{b}_t$ is a vector of the most disaggregated series, and $\bm{S}_t$ shows how the two are related.

\cite{HAA+11} considered four hierarchical forecasting methods for univariate time series, namely the top-down, bottom-up, middle-out and optimal combination methods. Among the four, only bottom-up and optimal combination methods are suitable for forecasting a non-unique group structure. These two methods are reviewed in Sections~\ref{sec:bu} and~\ref{sec:ols}, and their point and interval forecast accuracy comparisons with the independent forecasting method are presented in Sections~\ref{sec:point_compar} and~\ref{sec:interval_compar}, respectively.

\subsection{Bottom-up method}\label{sec:bu}

One of the commonly used methods to forecasting grouped time series is the bottom-up method \citep[e.g.,][]{DM92, ZT00}. This method involves first generating base forecasts for each of the most disaggregated series and then aggregating these to produce all required forecasts. For example, let us consider the Japanese data. We first generate $h$-step-ahead base forecasts for the most disaggregated series, namely $\widehat{\bm{b}}_{n+h} = \big[\widehat{R}_{\text{P1*F}, n+h}, \widehat{R}_{\text{P1*M},n+h}, \widehat{R}_{\text{P2*F}, n+h}, \widehat{R}_{\text{P2*M},n+h}, \dots,$
$\widehat{R}_{\text{P47*F},n+h}, \widehat{R}_{\text{P47*M},n+h}\big]^\top$.

Then the historical ratios that form the $\bm{S}_t$ summing matrix are forecast using an automated ARIMA algorithm \citep{HK08}. That is, let $p_t = E_{\text{X*S},t}/E_{\text{Y*W},t}$ be a non-zero element of $\bm{S}_t$. We forecast each time series $\{p_1,\dots,p_n\}$ for $h$-step-ahead to obtain $\hat{p}_{n+h}$. These are then used to form the matrix $\bm{S}_{n+h}$. Thus we obtain reconciled forecasts for all series:
\begin{equation*}
\overline{\bm{R}}_{n+h} = \bm{S}_{n+h} \widehat{\bm{b}}_{n+h}.
\end{equation*}

The bottom-up method has the agreeable feature that it is simple and intuitive, and always results in series that are ``aggregate consistent'' (i.e., that the resulting forecasts satisfy the same aggregation constraints as the original data). The method performs well when the signal-to-noise ratio is relatively strong for the most disaggregated series. On the other hand, it may lead to inaccurate forecasts of the top-level series, in particular when there are missing or noisy data at the bottom level \citep[see for example,][in the univariate time series context]{SW79,STM88}.

\subsection{Optimal combination method}\label{sec:ols}

Instead of considering only the bottom-level series, \citet{HAA+11} proposed a method in which base forecasts for all aggregated and disaggregated series are computed independently, and then the resulting forecasts are reconciled so that they satisfy the aggregation constraints.  As the base forecasts are independently generated, they will not usually be ``aggregate consistent''. The optimal combination method combines the base forecasts through linear regression by generating a set of revised forecasts that are as close as possible to the base forecasts but that also aggregate consistently within the group. The method is derived by writing the base forecasts as the response variable of the linear regression
\begin{equation*}
\widehat{\bm{R}}_{n+h} = \bm{S}_{n+h} \bm{\beta}_{n+h} + \bm{\varepsilon}_{n+h},
\end{equation*}
where $\widehat{\bm{R}}_{n+h}$ is a matrix of $h$-step-ahead base forecasts for all series, stacked in the same order as for original data; $\bm{\beta}_{n+h} = \text{E}[\bm{b}_{n+h}\mid \bm{R}_1,\dots,\bm{R}_n]$ is the unknown mean of the forecast distributions of the most disaggregated series; and $\bm{\varepsilon}_{n+h}$ represents the reconciliation errors.

To estimate the regression coefficients, \cite{HAA+11} and \citet{Hyndman2016-wp} proposed a weighted least squares solution which we adapt to our problem as follows:
\begin{equation*}
\widehat{\bm{\beta}}_{n+h} = \left(\bm{S}_{n+h}^{\top}\bm{W}^{-1}\bm{S}_{n+h}\right)^{-1}\bm{S}_{n+h}^{\top} \bm{W}^{-1} \widehat{\bm{R}}_{n+h},
\end{equation*}
where $\bm{W}$ is a diagonal matrix containing the one-step-ahead forecast variances for each series. Then the revised forecasts are given by
\begin{equation*}
\overline{\bm{R}}_{n+h} = \bm{S}_{n+h} \widehat{\bm{\beta}}_{n+h}
 = \bm{S}_{n+h} \left(\bm{S}_{n+h}^{\top}\bm{S}_{n+h}\right)^{-1}\bm{S}_{n+h}^{\top} \widehat{\bm{R}}_{n+h}.
\end{equation*}
By construction, these are aggregate consistent and involve a combination of all the base forecasts. They are also unbiased since
$\E[\overline{\bm{R}}_{n+h}] = \bm{S}_{n+h} \bm{\beta}_{n+h}$.

\subsection{Constructing uniform and pointwise prediction intervals}

To assess the forecast uncertainty, we adapt the method of \cite{ANH15} for computing uniform and pointwise prediction intervals. The method can be summarized in the following steps:
\begin{enumerate}
\item Using all observed data, compute the $K$-variate score vectors $(\bm{\beta}_1,\dots,\bm{\beta}_{K})$ and the sample functional principal components $\left[\widehat{\phi}_1(z),\dots,\widehat{\phi}_K(z)\right]$. Then, we can construct in-sample forecasts
\begin{equation*}
\F_{\zeta+h}(z) = \widehat{\beta}_{\zeta+h,1}\widehat{\phi}_1(z)+ \cdots + \widehat{\beta}_{\zeta+h,K}\widehat{\phi}_K(z),
\end{equation*}
where $(\widehat{\beta}_{\zeta+h,1},\dots,\widehat{\beta}_{\zeta+h,K})$ are the elements of the $h$-step-ahead prediction obtained from $(\bm{\beta}_1,\dots,\bm{\beta}_K)$ by a means of univariate time-series forecasting method, for $\zeta\in \{K,\dots,n-h\}$.
\item With the in-sample forecasts, we calculate the in-sample forecast errors
\begin{equation*}
\widehat{\epsilon}_{\omega}(z) = \F_{\zeta+h}(z) - \widehat{\F}_{\zeta+h}(z),
\end{equation*}
where $\omega \in \{1, 2,\dots,M\}$ and $M = n-h-K+1$.
\item Based on these in-sample forecast errors, we can sample with replacement to obtain a series of bootstrapped forecast errors, from which we obtain lower and upper bounds, denoted by $\gamma^{l}(z)$ and $\gamma^{u}(z)$, respectively. We then seek a tuning parameter $\varphi_{\alpha}$ such that $\alpha\times 100\%$ of the residual functions satisfy
\begin{equation*}
\varphi_{\alpha}\times \gamma^{l}(z)\leq \widehat{\epsilon}_{\omega}(z)\leq \varphi_{\alpha}\times \gamma^{u}(z),\qquad z\in \mathcal{I}.
\end{equation*}
The residuals $\widehat{\epsilon}_1(z),\dots,\widehat{\epsilon}_M(z)$ are expected to be approximately stationary and, by the law of large numbers, to satisfy
\begin{multline*}
\frac{1}{M}\sum^M_{\omega=1} \mathbb{1}\left(\varphi_{\alpha}\times \gamma^{l}(z)\leq \widehat{\epsilon}_{\omega}(z)\leq \varphi_{\alpha}\times \gamma^{u}(z)\right) \\
\approx \text{Pr}\left[\varphi_{\alpha}\times \gamma^{l}(z) \leq \F_{n+h}(z) - \widehat{\F}_{n+h}(z)\leq \varphi_{\alpha}\times \gamma^{u}(z)\right].
\end{multline*}
\end{enumerate}

Note that \cite{ANH15} calculate the standard deviation of $\left[\widehat{\epsilon}_{1}(z), \dots, \widehat{\epsilon}_{M}(z)\right]$, which leads to a parametric approach of constructing prediction intervals. Here we consider a nonparametric approach, as it allows us to reconcile bootstrapped forecasts among different functional time series in a hierarchy. Step 3 can easily be extended to pointwise prediction interval, where we determine a tuning parameter $\pi_{\alpha}$ such that $\alpha\times 100\%$ of the residual data points satisfy
\begin{equation*}
\pi_{\alpha}\times \gamma^{l}(z_j) \leq \widehat{\epsilon}_{\omega}(z_j) \leq \pi_{\alpha}\times \gamma^{u}(z_j),
\end{equation*}
where $j$ symbolizes discretized data points. Then, the $h$-step-ahead pointwise prediction intervals are given as
\begin{equation*}
\pi_{\alpha}\times \gamma^{l}(z_j) \leq \F_{n+h}(z_j) - \widehat{\F}_{n+h}(z_j) \leq \pi_{\alpha}\times \gamma^{u}(z_j).
\end{equation*}

\section{Results of the point forecasts}\label{sec:5}

\subsection{Point forecast evaluation}

An expanding window analysis of a time series model is commonly used to assess model and parameter stabilities over time. It assesses the constancy of a model's parameter by computing parameter estimates and their forecasts over an expanding window of a fixed size through the sample \citep[see][Chapter 9 for details]{ZW06}. Using the first 29 observations from 1975 to 2003 in the Japanese age-specific mortality rates, we produce  one- to ten-step-ahead point forecasts. Through an expanding window approach, we re-estimate the parameters in the univariate time series forecasting models using the first 30 observations from 1975 to 2004. Forecasts from the estimated models are then produced for one to nine-step-ahead. We iterate this process by increasing the sample size by one year until reaching the end of data period in 2013. This process produces 10 one-step-ahead forecasts, 9 two-step-ahead forecasts, \dots, and 1 ten-step-ahead forecast. We compare these forecasts with the holdout samples to determine the out-of-sample point forecast accuracy.

To evaluate the point forecast accuracy, we use the mean absolute forecast error (MAFE) and root mean squared forecast error (RMSFE). They measure how close the forecasts are in comparison to the actual values of the variable being forecast. For each series $k$, and they can be written as
\begin{align*}
\text{MAFE}_k(h) &= \frac{1}{101\times (11-h)}\sum^{10}_{\varsigma=h}\sum^{101}_{j=1}\left|\F_{n+\varsigma}^k(z_j) - \widehat{\F}_{n+\varsigma}^k(z_j)\right|, \\
\text{RMSFE}_k(h) &= \sqrt{\frac{1}{101\times (11-h)}\sum^{10}_{\varsigma=h}\sum^{101}_{j=1}\left[\F_{n+\varsigma}^k(z_j) - \widehat{\F}_{n+\varsigma}^k(z_j)\right]^2},
\end{align*}
where $\F_{n+\varsigma}^k(z_j)$ represents the actual holdout sample for the $j$th age and $\varsigma$th curve of the forecasting period in the $k$th series, while $\widehat{\F}_{n+\varsigma}^k(z_j)$ represents the point forecasts for the holdout sample.

By averaging MAFE$_k(h)$ and RMSFE$_k(h)$ across the number of series within each level of disaggregation, we obtain an overall assessment of the point forecast accuracy for each level within the collection of series, denoted by MAFE$(h)$ and RMSFE$(h)$. They are defined as
\begin{align*}
\text{MAFE}(h) = \frac{1}{m_k}\sum^{m_k}_{k=1}\text{MAFE}_k(h), \qquad
\text{RMSFE}(h) = \frac{1}{m_k}\sum^{m_k}_{k=1}\text{RMSFE}_k(h),
\end{align*}
where $m_k$ denotes the number of series at the $k$th level of disaggregation, for $k=1,\dots,K$.

For 10 different forecast horizons, we consider two summary statistics to evaluate point forecast accuracy between the methods for national and sub-national population. The summary statistics chosen are the mean and median values due to their suitability for handling squared and absolute errors \citep{Gneiting11}. They are given by
\begin{align*}
\text{Mean}\left(\text{RMSFE}\right) = \frac{1}{10}\sum^{10}_{h=1}\text{RMSFE}(h), \qquad
 \text{Median}\left(\text{MAFE}\right) = \frac{1}{2}\left[\text{MAFE}(5) + \text{MAFE}(6)\right],
\end{align*}
where ${[5]}$ and ${[6]}$ represent the $5$th and $6$th terms after ranking $\text{MAFE}(h)$ for $h=1,2,\dots,10$ from smallest to largest.

\subsection{Point forecast comparison}\label{sec:point_compar}

Averaging over all series at each level of the Japanese data hierarchy, Tables~\ref{tab:mae} and~\ref{tab:rmse} present MAFE$(h)$ and RMSFE$(h)$ values using the independent functional time series and two grouped functional time series forecasting methods. The bold entries highlight the method that performs the best for each level of the hierarchy and each forecast horizon, based on the smallest forecast error. In the short-term forecast horizon, the independent functional time series forecasting and optimal combination methods generally have the smaller forecast errors than the bottom-up method. As the forecast horizon increases from $h=3$ to $h=10$, the bottom-up method performs the best with the smallest forecast errors. At the bottom level, it is not surprising that the independent functional time series and bottom-up methods produce the same forecast accuracy. Averaged over all levels of a hierarchy, it is advantageous to use the grouped functional time series forecasting methods over the independent functional time series forecasting method. For this example, we recommend the bottom-up method.

\begin{table}[!htbp]
\caption{MAFEs ($\times 100$) in the holdout sample between the independent functional time series forecasting and two grouped functional time series forecasting methods applied to the Japanese age-specific mortality rates. The bold entries highlight the method that performs best for each level of the hierarchy and each forecast horizon, as well as summary statistic.}\label{tab:mae}
\tabcolsep 0.1in\spacingset{1}\small\centering
\begin{tabular}{@{\extracolsep{4pt}} llcccccc@{}}
  \\\toprule
Forecasting  & $h$ & Total & Sex & Region & Region & Prefecture & Prefecture   \\
method		& & & & &  (Sex) & & (Sex) \\
\midrule
Independent & 1 & $0.134$ & $0.133$ & $\textBF{0.157}$ & $0.209$ & $0.252$ & $0.378$ \\
& 2 & $0.194$ & $0.181$ & $0.189$ & $\textBF{0.225}$ & $\textBF{0.253}$ & $0.390$ \\
& 3 & $0.220$ & $0.213$ & $0.212$ & $0.235$ & $0.263$ & $\textBF{0.365}$ \\
& 4 & $0.256$ & $0.259$ & $0.248$ & $0.262$ & $0.279$ & $\textBF{0.374}$ \\
& 5 & $0.290$ & $0.301$ & $0.272$ & $0.287$ & $0.292$ & $\textBF{0.381}$ \\
& 6 & $0.323$ & $0.334$ & $0.300$ & $0.312$ & $0.311$ & $\textBF{0.399}$ \\
& 7 & $0.375$ & $0.393$ & $0.347$ & $0.357$ & $0.337$ & $\textBF{0.420}$ \\
& 8 & $0.415$ & $0.432$ & $0.388$ & $0.398$ & $0.367$ & $\textBF{0.445}$ \\
& 9 & $0.461$ & $0.460$ & $0.412$ & $0.411$ & $0.378$ & $\textBF{0.451}$ \\
& 10 & $0.457$ & $0.427$ & $0.395$ & $0.391$ & $0.366$ & $\textBF{0.437}$ \\\cmidrule{2-8}
& Median & $0.306$ & $0.318$ & $0.286$ & $0.299$ & $0.301$ & $\textBF{0.394}$ \\ \midrule
Bottom-up & 1 & $0.116$ & $0.134$ & $0.179$ & $0.220$ & $0.256$ & $0.378$ \\
& 2 & $0.123$ & $\textBF{0.142}$ & $0.196$ & $0.235$ & $0.273$ & $0.390$ \\
& 3 & $\textBF{0.129}$ & $\textBF{0.151}$ & $\textBF{0.166}$ & $\textBF{0.216}$ & $\textBF{0.242}$ & $\textBF{0.365}$ \\
& 4 & $\textBF{0.142}$ & $\textBF{0.178}$ & $\textBF{0.177}$ & $\textBF{0.234}$ & $\textBF{0.248}$ & $\textBF{0.374}$ \\
& 5 & $\textBF{0.138}$ & $\textBF{0.202}$ & $\textBF{0.178}$ & $\textBF{0.249}$ & $\textBF{0.249}$ & $\textBF{0.381}$ \\
& 6 & $\textBF{0.160}$ & $\textBF{0.234}$ & $\textBF{0.192}$ & $\textBF{0.273}$ & $\textBF{0.260}$ & $\textBF{0.399}$ \\
& 7 & $\textBF{0.179}$ & $\textBF{0.283}$ & $\textBF{0.211}$ & $\textBF{0.313}$ & $\textBF{0.268}$ & $\textBF{0.420}$ \\
& 8 & $\textBF{0.205}$ & $\textBF{0.322}$ & $\textBF{0.236}$ & $\textBF{0.354}$ & $\textBF{0.283}$ & $\textBF{0.445}$ \\
& 9 & $\textBF{0.228}$ & $\textBF{0.353}$ & $\textBF{0.248}$ & $\textBF{0.371}$ & $\textBF{0.283}$ & $\textBF{0.451}$ \\
& 10 & $\textBF{0.209}$ & $\textBF{0.329}$ & $\textBF{0.231}$ & $\textBF{0.354}$ & $\textBF{0.267}$ & $\textBF{0.437}$ \\\cmidrule{2-8}
& Median & $\textBF{0.151}$ & $\textBF{0.218}$ & $\textBF{0.194}$ & $\textBF{0.261}$ & $\textBF{0.264}$ & $\textBF{0.394}$ \\\hline
Optimal combination & 1 & $\textBF{0.111}$ & $\textBF{0.130}$ & $0.164$ & $\textBF{0.207}$ & $\textBF{0.247}$ & $\textBF{0.371}$ \\
& 2 & $\textBF{0.120}$ & $0.149$ & $\textBF{0.181}$ & $0.226$ & $0.261$ & $\textBF{0.383}$ \\
& 3 & $0.139$ & $0.176$ & $0.168$ & $0.224$ & $0.246$ & $0.373$ \\
& 4 & $0.164$ & $0.217$ & $0.190$ & $0.255$ & $0.258$ & $0.388$ \\
& 5 & $0.183$ & $0.258$ & $0.203$ & $0.284$ & $0.266$ & $0.404$ \\
& 6 & $0.208$ & $0.293$ & $0.223$ & $0.314$ & $0.280$ & $0.426$ \\
& 7 & $0.248$ & $0.352$ & $0.255$ & $0.364$ & $0.299$ & $0.456$ \\
& 8 & $0.280$ & $0.394$ & $0.291$ & $0.413$ & $0.321$ & $0.487$ \\
& 9 & $0.301$ & $0.422$ & $0.302$ & $0.427$ & $0.326$ & $0.497$ \\
& 10 & $0.282$ & $0.399$ & $0.286$ & $0.412$ & $0.310$ & $0.483$ \\\cmidrule{2-8}
& Median & $0.195$ & $0.276$ & $0.213$ & $0.299$ & $0.273$ & $0.415$ \\
\bottomrule
\end{tabular}
\end{table}

\begin{table}[!htbp]
 \caption{RMSFEs ($\times 100$) in the holdout sample between the independent functional time series forecasting and two grouped functional time series forecasting methods applied to the Japanese age-specific mortality rates. The bold entries highlight the method that performs best for each level of the hierarchy and each forecast horizon, as well as summary statistic.}\label{tab:rmse} 
\tabcolsep 0.1in\spacingset{1}\small\centering
 \begin{tabular}{@{\extracolsep{5pt}} llcccccc@{}}
\\\toprule
Forecasting  & $h$ & Total & Sex & Region & Region & Prefecture & Prefecture   \\
method		& & & & &  (Sex) & & (Sex) \\
\midrule
Independent & 1 & $0.468$ & $\textBF{0.464}$ & $\textBF{0.528}$ & $0.719$ & $\textBF{0.812}$ & $1.300$ \\
& 2 & $0.589$ & $0.573$ & $\textBF{0.611}$ & $\textBF{0.765}$ & $\textBF{0.814}$ & $1.367$ \\
& 3 & $0.658$ & $0.657$ & $0.680$ & $0.804$ & $0.843$ & $\textBF{1.235}$ \\
& 4 & $0.740$ & $0.776$ & $0.765$ & $0.880$ & $0.885$ & $\textBF{1.264}$ \\
& 5 & $0.812$ & $0.876$ & $0.824$ & $0.951$ & $0.917$ & $\textBF{1.285}$ \\
& 6 & $0.876$ & $0.946$ & $0.876$ & $0.996$ & $0.958$ & $\textBF{1.320}$ \\
& 7 & $0.992$ & $1.087$ & $0.982$ & $1.117$ & $1.021$ & $\textBF{1.375}$ \\
& 8 & $1.084$ & $1.176$ & $1.068$ & $1.205$ & $1.077$ & $\textBF{1.418}$ \\
& 9 & $1.170$ & $1.222$ & $1.101$ & $1.210$ & $1.084$ & $\textBF{1.399}$ \\
& 10 & $1.135$ & $1.107$ & $1.042$ & $1.127$ & $1.024$ & $\textBF{1.331}$ \\\cmidrule{2-8}
& Mean & $0.852$ & $0.888$ & $0.848$ & $0.977$ & $0.943$ & $\textBF{1.330}$ \\
\midrule
Bottom up & 1 & $\textBF{0.413}$ & $0.469$ & $0.614$ & $0.740$ & $0.856$ & $1.300$ \\
& 2 & $\textBF{0.423}$ & $\textBF{0.495}$ & $0.729$ & $0.836$ & $0.956$ & $1.367$ \\
& 3 & $\textBF{0.466}$ & $\textBF{0.549}$ & $\textBF{0.570}$ & $\textBF{0.742}$ & $\textBF{0.778}$ & $\textBF{1.235}$ \\
& 4 & $\textBF{0.513}$ & $\textBF{0.624}$ & $\textBF{0.613}$ & $\textBF{0.800}$ & $\textBF{0.804}$ & $\textBF{1.264}$ \\
& 5 & $\textBF{0.540}$ & $\textBF{0.692}$ & $\textBF{0.637}$ & $\textBF{0.854}$ & $\textBF{0.812}$ & $\textBF{1.285}$ \\
& 6 & $\textBF{0.579}$ & $\textBF{0.750}$ & $\textBF{0.671}$ & $\textBF{0.900}$ & $\textBF{0.840}$ & $\textBF{1.320}$ \\
& 7 & $\textBF{0.643}$ & $\textBF{0.865}$ & $\textBF{0.736}$ & $\textBF{1.011}$ & $\textBF{0.875}$ & $\textBF{1.375}$ \\
& 8 & $\textBF{0.706}$ & $\textBF{0.948}$ & $\textBF{0.794}$ & $\textBF{1.099}$ & $\textBF{0.910}$ & $\textBF{1.418}$ \\
& 9 & $\textBF{0.744}$ & $\textBF{1.000}$ & $\textBF{0.815}$ & $\textBF{1.116}$ & $\textBF{0.907}$ & $\textBF{1.399}$ \\
& 10 & $\textBF{0.673}$ & $\textBF{0.899}$ & $\textBF{0.752}$ & $\textBF{1.038}$ & $\textBF{0.842}$ & $\textBF{1.331}$ \\\cmidrule{2-8}
& Mean & $\textBF{0.570}$ & $\textBF{0.729}$ & $\textBF{0.693}$ & $\textBF{0.914}$ & $\textBF{0.858}$ & $\textBF{1.330}$ \\
\midrule
Optimal combination & 1 & $0.430$ & $0.490$ & $0.571$ & $\textBF{0.712}$ & $0.816$ & $\textBF{1.276}$ \\
& 2 & $0.462$ & $0.546$ & $0.654$ & $0.795$ & $0.881$ & $\textBF{1.327}$ \\
& 3 & $0.527$ & $0.619$ & $0.606$ & $0.782$ & $0.805$ & $1.265$ \\
& 4 & $0.592$ & $0.714$ & $0.666$ & $0.863$ & $0.843$ & $1.307$ \\
& 5 & $0.644$ & $0.805$ & $0.710$ & $0.939$ & $0.867$ & $1.343$ \\
& 6 & $0.694$ & $0.875$ & $0.754$ & $0.996$ & $0.901$ & $1.387$ \\
& 7 & $0.779$ & $1.004$ & $0.839$ & $1.122$ & $0.957$ & $1.459$ \\
& 8 & $0.851$ & $1.094$ & $0.913$ & $1.220$ & $1.004$ & $1.513$ \\
& 9 & $0.889$ & $1.146$ & $0.926$ & $1.234$ & $1.003$ & $1.497$ \\
& 10 & $0.819$ & $1.048$ & $0.865$ & $1.155$ & $0.936$ & $1.427$ \\\cmidrule{2-8}
& Mean & $0.669$ & $0.834$ & $0.750$ & $0.982$ & $0.901$ & $1.380$ \\
\bottomrule
\end{tabular}
\end{table}

\subsection{Comparison with moving functional median}

As a comparison, we consider a moving functional median method to produce point forecasts. The functional median allows us to rank a sample of curves based on their location depth; i.e., the distance from the functional median (the deepest curve). This leads to the notion of functional depth \citep[see, e.g.,][]{CFF06,CFF07}. 

We briefly describe one functional depth measure, namely \citeauthor{FM01}'s \citeyearpar{FM01} depth. For each $z\in \mathcal{I}$, let $F_{n,z}$ be the empirical sample distribution of $\{\mathcal{X}_1(z),\mathcal{X}_2(z),\dots,\mathcal{X}_n(z)\}$ and let $Z_i(z)$ be the univariate depth of function $\mathcal{X}_i(z)$, given by
\begin{equation*}
I_i = \int_{\mathcal{I}}Z_i(z)dz = \int_{\mathcal{I}} 1- \Big|\frac{1}{2} - F_{n,z}(\mathcal{X}_i(z))\Big|dz,
\end{equation*}
and the values of $I_i$ provide a way of ranking curves from inward to outward. Thus, the functional median is the deepest curve with the maximum $I_i$ value.

Using the expanding window approach, we compute the moving functional median and report one to ten-step-ahead point forecast accuracy in Table~\ref{tab:functional_median}. Since all the functional time series considered are non-stationary in nature, the functional median cannot rapidly capture the dynamic changes in the underlying patterns, and thus does not perform as well as the proposed functional time series method.

\begin{table}[!htbp]
\centering
\caption{MAFE and RMSFE in the holdout sample using the moving functional median method applied to the Japanese age-specific mortality rates.}\label{tab:functional_median}
\tabcolsep 0.13in\spacingset{1}\small\centering
\begin{tabular}{@{}llrrrrrr@{}}
\toprule
Error & $h$ & Total  & Sex    & Region & Region + Sex & Prefecture & Prefecture + Sex \\
\midrule
MAFE  & 1   & 0.0685 & 0.0736 & 0.0688 & 0.0738       & 0.0695     & 0.0755 \\
      & 2   & 0.0676 & 0.0727 & 0.0679 & 0.0729       & 0.0686     & 0.0746 \\
      & 3   & 0.0665 & 0.0714 & 0.0669 & 0.0717       & 0.0675     & 0.0734 \\
      & 4   & 0.0656 & 0.0703 & 0.0659 & 0.0706       & 0.0665     & 0.0723 \\
      & 5   & 0.0647 & 0.0693 & 0.0650 & 0.0695       & 0.0655     & 0.0712 \\
      & 6   & 0.0638 & 0.0683 & 0.0640 & 0.0684       & 0.0645     & 0.0701 \\
      & 7   & 0.0630 & 0.0675 & 0.0633 & 0.0676       & 0.0637     & 0.0691 \\
      & 8   & 0.0623 & 0.0664 & 0.0624 & 0.0667       & 0.0629     & 0.0682 \\
      & 9   & 0.0618 & 0.0657 & 0.0623 & 0.0659       & 0.0627     & 0.0678 \\
      & 10  & 0.0627 & 0.0669 & 0.0638 & 0.0673       & 0.0643     & 0.0691 \\
      \cmidrule{2-8}
      & Median & 0.0642 & 0.0688 & 0.0645 & 0.0690 & 0.0650 & 0.0706 \\
\midrule
RMSFE & 1   & 0.1234 & 0.1318 & 0.1243 & 0.1329       & 0.1265     & 0.1383 \\
      & 2   & 0.1222 & 0.1306 & 0.1232 & 0.1316       & 0.1255     & 0.1371 \\
      & 3   & 0.1207 & 0.1289 & 0.1218 & 0.1300       & 0.1240     & 0.1355 \\
      & 4   & 0.1193 & 0.1273 & 0.1204 & 0.1283       & 0.1224     & 0.1339 \\
      & 5   & 0.1179 & 0.1258 & 0.1190 & 0.1267       & 0.1208     & 0.1321 \\
      & 6   & 0.1166 & 0.1241 & 0.1175 & 0.1248       & 0.1190     & 0.1301 \\
      & 7   & 0.1152 & 0.1227 & 0.1162 & 0.1232       & 0.1175     & 0.1281 \\
      & 8   & 0.1140 & 0.1209 & 0.1144 & 0.1214       & 0.1159     & 0.1260 \\
      & 9   & 0.1125 & 0.1193 & 0.1138 & 0.1197       & 0.1150     & 0.1245 \\
      & 10  & 0.1136 & 0.1209 & 0.1163 & 0.1218       & 0.1177     & 0.1267 \\
      \cmidrule{2-8}
      & Mean & 0.1170 & 0.1255 & 0.1175 & 0.1267 & 0.1208 & 0.1320 \\
\bottomrule
\end{tabular}
\end{table}

\section{Results of the interval forecasts}\label{sec:7}

\subsection{Interval forecast evaluation}

In order to evaluate pointwise interval forecast accuracy, we utilize the interval score of \cite{GR07} \citep[see also][]{GK14}. For each year in the forecasting period, the $h$-step-ahead prediction intervals were calculated at the $100(1-\alpha)\%$ nominal coverage probability. We consider the common case of the symmetric $100(1-\alpha)\%$ prediction interval, with lower and upper bounds that are predictive quantiles at $\alpha/2$ and $1-\alpha/2$, denoted by $\widehat{\mathcal{X}}_{n+h}^{l}(z_j)$ and $\widehat{\mathcal{X}}_{n+h}^{u}(z_j)$. As defined by \cite{GR07}, a scoring rule for the pointwise interval forecast at time point $z_j$ is
\begin{align*}
S_{\alpha}\left[\widehat{\F}_{n+h}^l(z_j),\widehat{\F}_{n+h}^u(z_j);\F_{n+h}(z_j)\right] = & \left[\widehat{\F}_{n+h}^u(z_j) - \widehat{\F}_{n+h}^l(z_j)\right] + \\
&\frac{2}{\alpha}\left[\widehat{\F}_{n+h}^l(z_j)-\F_{n+h}(z_j)\right]\mathbb{1}\left\{\F_{n+h}(z_j)<\widehat{\F}_{n+h}^l(z_j)\right\} + \\
&\frac{2}{\alpha}\left[\F_{n+h}(z_j)-\widehat{\F}^u_{n+h}(z_j)\right]\mathbb{1}\left\{\F_{n+h}(z_j)>\widehat{\F}_{n+h}^u(z_j)\right\},
\end{align*}
where $\alpha$ denotes the level of significance, customarily $\alpha=0.2$. The interval score rewards a narrow prediction interval, if and only if the true observation lies within the prediction interval. The optimal interval score is achieved when $\F_{n+h}(z_j)$ lies between $\widehat{\F}^l_{n+h}(z_j)$ and $\widehat{\F}^u_{n+h}(z_j)$, and the distance between $\widehat{\F}_{n+h}^l(z_j)$ and $\widehat{\F}_{n+h}^u(z_j)$ is minimal.

For different time points in a curve and different days in the forecasting period, the mean interval score is defined by
\begin{equation*}
\overline{S}_{\alpha}(h) = \frac{1}{101\times (11-h)}\sum^{10}_{\varsigma=h}\sum^{101}_{j=1}S_{\alpha}\left[\widehat{\F}_{n+\varsigma}^l(z_j),\widehat{\F}_{n+\varsigma}^u(z_j);\F_{n+\varsigma}(z_j)\right] ,
\end{equation*}
where $S_{\alpha}\left[\widehat{\F}_{n+\varsigma}^l(z_j),\widehat{\F}_{n+\varsigma}^u(z_j);\F_{n+\varsigma}(z_j)\right] $ denotes the interval score at the $\varsigma$th curve of the forecasting period.

For 10 different forecast horizons, we consider two summary statistics to evaluate interval forecast accuracy. The summary statistics chosen are the mean and median values, given by
\begin{align*}
\text{Mean}\left(\overline{S}_{\alpha}\right) = \frac{1}{10}\sum^{10}_{h=1}\overline{S}_{\alpha}(h), \qquad
\text{Median}\left(\overline{S}_{\alpha}\right) = \frac{1}{2}\left[\overline{S}_{\alpha}(5) + \overline{S}_{\alpha}(6)\right].
\end{align*}

\subsection{Interval forecast comparison}\label{sec:interval_compar}

\begin{table}[p]
  \caption{Mean interval scores ($\times 100$) in the holdout sample between the independent functional time series forecasting and two grouped functional time series forecasting methods applied to the Japanese age-specific mortality rates. The bold entries highlight the method that performs best for each level of the hierarchy and each forecast horizon, as well as two summary statistics.}\label{tab:3}
\tabcolsep 0.095in\spacingset{1}\small\centering
\begin{tabular}{@{\extracolsep{5pt}} llcccccc@{}}
\toprule
Forecasting  & $h$ & Total & Sex & Region & Region & Prefecture & Prefecture   \\
method		& & & & &  (Sex) & & (Sex) \\
\midrule
Independent & 1 & $\textBF{0.523}$ & $\textBF{0.627}$ & $1.396$ & $2.184$ & $\textBF{1.396}$ & $ 2.184$ \\
& 2 & $\textBF{0.676}$ & $\textBF{0.819}$ & $1.372$ & $2.276$ & $\textBF{1.372}$ & $2.276$ \\
& 3 & $\textBF{0.738}$ & $\textBF{0.914}$ & $1.373$ & $2.060$ & $\textBF{1.373}$ & $2.060$ \\
& 4 & $\textBF{0.910}$ & $\textBF{1.076}$ & $1.421$ & $2.091$ & $1.421$ & $2.091$ \\
& 5 & $\textBF{1.123}$ & $\textBF{1.249}$ & $1.484$ & $2.104$ & $1.484$ & $2.104$ \\
& 6 & $\textBF{1.198}$ & $\textBF{1.315}$ & $1.565$ & $2.175$ & $1.565$ & $2.175$ \\
& 7 & $\textBF{1.322}$ & $\textBF{1.557}$ & $1.643$ & $2.250$ & $1.643$ & $2.250$ \\
& 8 & $\textBF{1.390}$ & $\textBF{1.666}$ & $1.734$ & $2.350$ & $1.734$ & $2.350$ \\
& 9 & $\textBF{1.558}$ & $\textBF{1.720}$ & $1.754$ & $2.294$ & $1.754$ & $2.294$ \\
& 10 & $1.437$ & $\textBF{1.580}$ & $1.752$ & $2.307$ & $1.752$ & $2.307$ \\\cmidrule{2-8}
& Mean & $\textBF{1.088}$ & $\textBF{1.252}$ & $1.549$ & $2.209$ & $1.549$ & $2.209$ \\
& Median & $\textBF{1.160}$ & $\textBF{1.282}$ & $1.524$ & $2.217$ & $1.524$ & $2.217$ \\
\midrule
Bottom up & 1 & $0.856$ & $0.832$ & $\textBF{0.974}$ & $1.166$ & $1.439$ & $2.184$ \\
& 2 & $0.972$ & $0.955$ & $\textBF{1.156}$ & $1.321$ & $1.578$ & $2.276$ \\
& 3 & $1.060$ & $1.079$ & $\textBF{0.885}$ & $\textBF{1.131}$ & $1.291$ & $2.060$ \\
& 4 & $1.206$ & $1.268$ & $\textBF{0.962}$ & $\textBF{1.207}$ & $1.316$ & $2.091$ \\
& 5 & $1.248$ & $1.406$ & $\textBF{0.978}$ & $\textBF{1.261}$ & $1.334$ & $2.104$ \\
& 6 & $1.354$ & $1.584$ & $\textBF{1.056}$ & $\textBF{1.370}$ & $1.372$ & $2.175$ \\
& 7 & $1.454$ & $1.859$ & $\textBF{1.120}$ & $\textBF{1.508}$ & $\textBF{1.414}$ & $2.250$ \\
& 8 & $1.538$ & $2.067$ & $\textBF{1.194}$ & $\textBF{1.664}$ & $\textBF{1.471}$ & $2.350$ \\
& 9 & $1.616$ & $2.166$ & $\textBF{1.185}$ & $\textBF{1.698}$ & $\textBF{1.409}$ & $2.294$ \\
& 10 & $\textBF{1.384}$ & $1.881$ & $\textBF{1.056}$ & $\textBF{1.562}$ & $1.449$ & $2.307$ \\\cmidrule{2-8}
& Mean & $1.269$ & $1.510$ & $\textBF{1.057}$ & $\textBF{1.389}$ & $1.407$ & $2.209$ \\
& Median & $1.301$ & $1.495$ & $\textBF{1.056}$ & $\textBF{1.346}$ & $1.412$ & $2.217$ \\
\midrule
Optimal combination & 1 & $0.924$ & $0.861$ & $0.995$ & $\textBF{1.101}$ & $\textBF{1.268}$ & $\textBF{2.029}$ \\
& 2 & $1.037$ & $1.032$ & $1.157$ & $\textBF{1.247}$ & $\textBF{1.363}$ & $\textBF{2.089}$ \\
& 3 & $1.241$ & $1.270$ & $1.066$ & $1.208$ & $\textBF{1.205}$ & $\textBF{1.968}$ \\
& 4 & $1.450$ & $1.560$ & $1.208$ & $1.354$ & $\textBF{1.251}$ & $\textBF{2.014}$ \\
& 5 & $1.582$ & $1.809$ & $1.289$ & $1.494$ & $\textBF{1.289}$ & $\textBF{2.054}$ \\
& 6 & $1.759$ & $2.063$ & $1.401$ & $1.656$ & $\textBF{1.344}$ & $\textBF{2.122}$ \\
& 7 & $1.957$ & $2.441$ & $1.556$ & $1.889$ & $1.416$ & $\textBF{2.208}$ \\
& 8 & $2.108$ & $2.712$ & $1.705$ & $2.107$ & $1.493$ & $\textBF{2.315}$ \\
& 9 & $2.207$ & $2.846$ & $1.719$ & $2.166$ & $1.468$ & $\textBF{2.284}$ \\
& 10 & $1.879$ & $2.490$ & $1.533$ & $1.981$ & $\textBF{1.429}$ & $\textBF{2.299}$ \\ \cmidrule{2-8}
& Mean & $1.614$ & $1.908$ & $1.363$ & $1.620$ & $\textBF{1.353}$ & $\textBF{2.138}$ \\
& Median & $1.670$ & $1.936$ & $1.345$ & $1.575$ & $\textBF{1.354}$ & $\textBF{2.105}$ \\
\bottomrule
\end{tabular}
\end{table}

In Table~\ref{tab:3}, we present the mean interval scores for one-step-ahead to ten-step-ahead forecasts, using the independent and two grouped functional time series forecasting methods. The independent functional time series generally gives the most accurate interval forecasts at the national level, while the grouped functional time series forecasting methods demonstrate superior forecast accuracy for the sub-national level. The bottom-up method gives the most accurate interval forecasts at the region level, while the optimal combination method gives the most accurate interval forecasts at the prefecture level. Based on the overall mean interval scores, the bottom-up methods outperform the independent functional time series forecasting and optimal combination methods, in terms of interval forecast accuracy. Thus, the bottom-up method is recommended for this example.

\section{Conclusion}\label{sec:conclu}

We have extended two grouped time series forecasting methods, namely the bottom-up and optimal combination methods, from univariate to functional time series. These grouped functional time series forecasting methods were derived by coupling grouped univariate time series forecasting methods with functional time series analysis.

The bottom-up method models and forecasts data series at the most disaggregated level, and then aggregates the results using the summing matrix. In that summing matrix, each element is forecast from the historical data using univariate time series models.

The optimal combination method combines the base forecasts obtained from independent functional time series forecasting methods using linear regression. It generates a set of revised forecasts that are as close as possible to the base forecasts, but that also aggregates consistently with the known grouping structure. Under some mild assumptions, the regression coefficient can be estimated by ordinary least squares.

Using age-specific mortality rates at the national and sub-national levels in Japan, we compare the one-step-ahead to ten-step-ahead forecast accuracy between the independent functional time series forecasting method and the two proposed grouped functional time series forecasting methods. We found that the grouped functional time series forecasting methods produced more accurate point and interval forecasts than those obtained by the independent functional time series forecasting method. In addition, the grouped functional time series forecasting methods produce forecasts that obey the natural group structure, thus giving forecast mortality rates at the sub-national levels that add up to the forecast mortality rates at the national level.

We have also presented a way of constructing uniform and pointwise prediction intervals for grouped functional time series using bootstrapping. The method calculates in-sample forecast errors between the in-sample holdout data and their reconstruction by functional principal component regression. By sampling with replacement from the bootstrapped in-sample errors, we obtain lower and upper bounds, and then find an optimal tuning parameter for achieving uniform or pointwise nominal coverage probability. With this tuning parameter, out-of-sample uniform or pointwise prediction intervals are obtained.

There are a few ways in which the paper can be further extended and we briefly outline four. 
First, due to the ready availability of suitable data, we have considered disaggregation of mortality by sex and geography. However, mortality rates can be further disaggregated with the inclusion of other factors, such as cause-of-death considered in \cite{ML97}, \cite{GK08} and \cite{GS15}, and socioeconomic status \textit{inter alia} \citep{BBA02, SAS+13}. With the appropriate data, it would be straightforward to extend our approach to take into account these further disaggregation factors. Second, coherent forecasting methods can be used to jointly model and forecast age-specific mortality rates from two or more populations \citep[see for example,][]{LL05, HBY13}. These methods could also be applied in the functional data context to ensure that related populations have non-diverging forecasts. Third, the proposed methodology can be applied to other application areas. To give just one example, university performance is commonly measured by student completion rates. The university-wide completion rates observed over years can be be disaggregated by age, sex, faculty, domestic or international status, and other factors. These disaggregations give us a group structure to constrain the forecasts of the age-specific completion rates, and to measure the effect of factors that may contribute to completions. Finally, since the presence of outliers can seriously affect the modeling and forecasting of principal component scores, a robust functional principal component decomposition \citep[such as proposed in][]{Bali2011-jf} and robust time series methods \citep[such as proposed in][]{Gelper2010-bd} can be adapted to our grouped functional time series methods. We leave each of these potential extensions to future research.

\newpage
\bibliographystyle{agsm}

\bibliography{gfts}
\end{document}